\documentclass[aps,pre,superscriptaddress,nofootinbib]{revtex4-2}

\usepackage[utf8]{inputenc} 
\usepackage[T1]{fontenc}    
\usepackage{hyperref}       
\usepackage{url}            
\usepackage{booktabs}       
\usepackage{amssymb,amsmath,amsfonts, dsfont}       
\usepackage{nicefrac}       
\usepackage{graphicx}       
\graphicspath{{./}} 
\usepackage{caption,subcaption}
\usepackage{float} 
\makeatletter
\let\newfloat\newfloat@ltx
\makeatother

\usepackage{algorithm}
\usepackage{algpseudocode} 
\usepackage{physics}
\usepackage{xcolor}
\usepackage{bm}
\usepackage{comment}
\usepackage{filecontents}

\newcommand{\mycolor}[1]{\textcolor{black}{#1}} 
\newcommand{\mycolorsecond}[1]{\textcolor{black}{#1}} 

\newcommand{\bea}{\begin{eqnarray}}
\newcommand{\eea}{\end{eqnarray}}
\newcommand{\bs}{\vec{s}}
\newcommand{\bmm}{\vec{m}}
\newcommand{\bh}{\vec{h}}
\newcommand{\bn}{\vec{n}}
\newcommand{\bH}{\vec{H}}

\newcommand{\<}{\langle}
\renewcommand{\>}{\rangle}
\newcommand{\E}{\mathbb{E}}

\newcommand{\1}{\mathds{1}}

\begin{document}

\title{Analytical solution to Heisenberg spin glass models on sparse random graphs and their de Almeida–Thouless line}

\author{Luca Maria Del Bono}
\affiliation{Dipartimento di Fisica, Sapienza Università di Roma, Piazzale Aldo Moro 5, Rome 00185, Italy}

\author{Flavio Nicoletti}
\affiliation{Dipartimento di Fisica, Sapienza Università di Roma, Piazzale Aldo Moro 5, Rome 00185, Italy}

\author{Federico Ricci-Tersenghi}
\affiliation{Dipartimento di Fisica, Sapienza Università di Roma, Piazzale Aldo Moro 5, Rome 00185, Italy}
\affiliation{CNR-Nanotec, Rome unit and INFN, sezione di Roma1, Piazzale Aldo Moro 5, Rome 00185, Italy}

\begin{abstract}

Results regarding spin glass models are, to this day, mainly confined to models with \mycolor{discrete (usually Ising)} spins. Spin glass models with continuous spins exhibit interesting new physical behaviors related to the additional degrees of freedom, but have been primarily studied on fully connected topologies. Only recently some advancements have been made in the study of continuous models on sparse graphs.
In this work we partially fill this void by introducing a method to solve numerically the Belief Propagation equations for systems of Heisenberg spins on sparse random graphs via a discretization of the sphere. We introduce techniques to study the finite-temperature, finite-connectivity case as well as novel algorithms to deal with the zero-temperature and large connectivity limits. As an application, we locate the de Almeida-Thouless line for this class of models and the critical field at zero temperature, showing the full consistency of the methods presented. 
Beyond the specific results reported for Heisenberg models, the approaches presented in this paper have a much broader scope of application and pave the way to the solution of strongly disordered models with continuous variables.

\end{abstract}

\maketitle

\section{Introduction}
In the last four decades, starting from the pioneering work of Parisi \cite{parisi1979infinite, parisi1979toward, parisi1980sequence, parisi1980magnetic, parisi1980order}, the theory of spin glasses has become a benchmark for the study of disordered and complex systems \cite{beyond_40ys}. While the nature of the transition in finite dimensions is not yet fully understood, solvable mean-field models of spin glasses have proven to be a solid framework for obtaining reliable qualitative predictions.  
The most famous spin glass solvable model is arguably the Sherrington-Kirkpatrick (SK)  model \cite{PhysRevLett.35.1792}, 
a fully connected model with Ising spins and random couplings. The solution to the model presents a paramagnetic solution at high temperature or large external field, which becomes locally unstable on the so-called de Almeida-Thouless \mycolor{(dAT)} critical line \cite{de1978stability}. Below such a critical line, the very rich spin glass phase appears, well described by the Replica Symmetry Breaking (RSB) scheme proposed solution by Parisi \cite{beyond}.

Most initial efforts were devoted to the study of the simpler case of \mycolor{discrete (usually Ising)} spins. However, in the last decade, there has been a renewed interest in the study of versions of the SK model with vector spin variables of fixed norm \cite{sharma_and_young, franz2022delocalization, franz2022linear}.
These models present a richer physics, due to the continuous nature of the degrees of freedom: of particular interest are the cases of spins with \mycolor{$m=2$} components (XY spins) and  \mycolor{$m=3$} components (Heisenberg spins), due to their ability to model a wide range of phenomena in condensed matter.

While in the absence of a magnetic field the behavior of the SK model with vector spins does not change qualitatively at the transition \mycolor{with respect to the Ising case} \cite{de1978infinite}, the application of a uniform field generates another instability line, the Gabay-Toulouse (GT) line \cite{gabay1981coexistence, gabay1982symmetry}.
\mycolor{In the presence of a uniform field, the symmetry along the direction of the field is always broken. Thus, on the GT line, only the degrees of freedom transverse to the direction of the external field can undergo a spontaneous symmetry breaking, which manifests in a weak form of RSB \cite{note1}. Consequently, the dAT line, which lies below the GT line (as exemplified in Fig. 7 of \cite{lupo2018comparison} in the case of the XY model), becomes just a crossover line from a weak to a strong RSB \cite{cragg1982spin}.} 
Indeed, the critical behavior on the GT line of anisotropic spin glasses with \mycolor{$m$}-dimensional spins is equivalent to the one of an isotropic spin glass with \mycolor{$(m-1)$}-dimensional spins \cite{moore1982critical}.

To avoid the GT transition and make the dAT line a distinct boundary between the RS and RSB phases, it is necessary to apply a field whose direction changes randomly from site to site, so that the system is on average isotropic. \mycolor{Notice that in the Ising case there is no need to distinguish the two cases, because one can be mapped to the other by applying a gauge transformation.} 

The location of the dAT line in the case of fields with Gaussian components of variance $H^2$ has been computed for general \mycolor{$m$} and fully connected (FC) topologies in \cite{sharma_and_young}. At low fields, one has the behavior $H \propto (T_g-T)^{3/2}$ which coincides with that found in \cite{de1978stability} for Ising variables. The computation in \cite{sharma_and_young} reports a zero temperature transition at $H_c(T = 0) = \frac{1}{\sqrt{m-2}}$: it is interesting to notice that, for both \mycolor{$m = 1$} (Ising) and \mycolor{$m = 2$} (XY) the FC models do not have such a $T=0$ transition in a field, given that the critical dAT line diverges to infinity for $T \to 0$. On the other hand, for \mycolor{$m > 2$} the critical field $H_c(T = 0)$ is finite and the dAT line has no divergence. More importantly, the nature of the $T=0$ phase transition at a finite field $H_c(T = 0)$ can be studied analytically by computing the spectrum of the energy Hessian via the random matrix theory. In Ref.~\cite{franz2022delocalization} it was shown that soft modes undergo a delocalization transition.

A natural question to ask is to what extent predictions on the critical behavior of spin glasses coming from mean-field theories apply to spin glasses in finite dimensions. While the zero field critical behavior of the SK model seems to qualitatively apply also in finite dimensions \cite{banos2010nature, banos2011sample}, the finite field and low-temperature regimes are less understood. In \cite{baity2015soft} it was found that at zero temperature the spin glass susceptibility of the 3D Heisenberg spin glass is non-divergent for any external field $H>0$. In contrast, in the vector SK model, the spin glass susceptibility diverges at the zero-temperature critical field \cite{sharma_and_young}. It then follows that, if the transition in a field exists in finite dimensions, at $T=0$ it must be qualitatively different from the one described by mean-field theory.
The standard step to go beyond mean-field theories on FC networks is to consider spin glasses defined on sparse random graphs. In the thermodynamic limit, the model can be solved exactly using the Cavity Method \cite{beyond}. While sparse random graphs retain an infinite-dimensional topology, the sparsity of the interaction network makes them closer to finite-dimensional systems.

When considering spin glass models on sparse random graphs the literature is very diverse. While models with Ising or Potts variables have received more attention \cite{klein1979, VianaBray1985, P.Mottishaw_1987, MezardParisi2001, MezardParisi2003, Krzakala_2008, Parisi_2014, Parisi2017, Concetti2018, desantis2019computation, Perrupato2022}, results on vector spin glasses are very scarce \cite{coolen2005finitely, lupo2017approximating, lupo2018comparison, lupo2019random, nicoletti2023low}. In particular, in the seminal paper \cite{coolen2005finitely} the authors were able to compute the critical points for XY and Heisenberg spin glasses on Erdos-Renyi graphs as functions of the mean connectivity and temperature, but only for zero external fields.

The important effects induced by an external field on the critical behavior of these models were considered only very recently\mycolor{, and only for the XY case,} \cite{lupo2018comparison,lupo2019random} thanks to the numerical approximations introduced in \cite{lupo2017approximating}.
In Ref.~\cite{lupo2018comparison} the authors study both GT and dAT lines, by considering the cases of constant field and randomly oriented fields respectively. In both cases the zero-temperature critical field is finite \cite{lupo2018comparison}. Moreover, they have shown explicitly the freezing phenomenon that differentiates between the GT and dAT lines. Finally, it has also been established in the XY case that the dAT line appears to be more robust to perturbations of the random field. 

However, extending the results above, obtained for the XY model, to models with more components (e.g. Heisenberg spins) turned out to be very difficult. The bottleneck is the required discretization of the variables domain: in XY models variables live on a circle which can be easily discretized by the clock model \cite{lupo2017approximating}, but Heisenberg variables live on a sphere and the sphere discretization is a notoriously difficult problem.
Only recently we came up with a new discretization scheme of the sphere that produces a very uniform set of discrete points \cite{delebono2024uniform} and this can be used for our purposes.

In this work, we consider Heisenberg spin glass models on random regular graphs (RRG) with randomly oriented external fields of fixed norm $H$. By solving the cavity equations of the model through the Population Dynamics Algorithm (PDA) we measure the location of the dAT line in the $(T,H)$ plane\mycolor{, by checking the stability of the paramagnetic phase. We focus on the dAT line, and not on the GT line described above, for two reasons. The first is that the dAT line, interestingly, identifies the transition to a non-zero-field RSB phase without altering the (average) symmetry between the spins \cite{sharma_and_young}, whereas the uniform field required to study the GT line introduces a preferential direction. The second is that the dAT line has been found, as mentioned earlier, more stable than the GT line in the XY case \cite{lupo2018comparison}. We point out, however, that the methods described in this paper can be applied straightforwardly to the study of the GT line as well.}

Firstly, we consider the case with connectivity $Z=3$, which corresponds to the sparsest non-trivial random graph. 
As a preliminary step, we study the effect of the sphere discretization on the estimate of the dAT line.
The $T=0$ case is studied separately, as a different algorithm is needed in this special case.
Secondly, we characterize the low temperature and low field regimes analytically and numerically, for growing values of the connectivity, and compare our results with the FC mean-field limit $Z=\infty$.

\mycolor{This paper is organized as follows. In Sec.~\ref{sec:model} we introduce the Heisenberg spin glass models on random graphs. In Sec.~\ref{sec:methods} we introduce the methods used to compute the marginals for the model and to locate the dAT line. Specifically, in Sec.~\ref{sec:ft_eq} we introduce the Belief Propagation (BP) equations at finite temperature, while in Sec.~\ref{sec:ft_alg} we describe how to solve these equations numerically as well as describing an algorithm to find the dAT line; in Sec.~\ref{sec:zt} we describe the zero-temperature limit of the Belief Propagation equations, introducing the max-sum equations (Sec.~\ref{sec:zt_eq}) and the corresponding algorithm to solve them and find the zero-temperature critical field (Sec.~\ref{sec:zt_eq}; in Sec.~\ref{sec:large} we study the large-connectivity limit of the model, introducing the large-connectivity expansion of the Belief Propagation equations and a novel algorithm, the Asymmetric Stability Checking. In Sec.~\ref{sec:results} we present the numerical results for the finite-temperature, finite-connectivity case (Sec.~\ref{sec:res_ft}) and the zero-temperature (Sec.~\ref{sec:res_0T}) and large-connectivity (Sec.~\ref{sec:res_large}) limits. Finally, in Sec.~\ref{sec:conclusions} we summarize our results and draw our conclusions.}

\section{Model}\label{sec:model}

The model we study is defined by the following Hamiltonian
\begin{equation}
\label{eq:Model}
    \mathcal{H}[\underline{{\bf s}}]\,=\,- \sum_{(i,j) \in E} J_{ij} \vec{s}_i \cdot \vec{s}_j - \sum_{i\in V} \vec{H}_i \cdot \vec{s}_i
\end{equation}
where $\vec{s}_i\in \mathbb{R}^3$ are $N$ Heisenberg spins with norm $|\vec{s}_i|=\sqrt{3}$, the couplings are quenched random parameters $J_{ij}=\pm1$ with equal probability and the fields $\vec{H}_i \in \mathbb{R}^3$ are quenched random parameters distributed uniformly on the sphere of radius $H$. 
These disorder distributions ensure a spatially homogeneous strength of the disorder, as $|J_{ij}|=1$ and $|\vec{H}_i|=H$ for any $(i,j)\in E$ and $i\in V$, being $E$ and $V$ the edge set and the vertex set of the random graph respectively.
The model is studied on random regular graphs (RRG) of size $|V|=N$ and fixed connectivity $Z$. The case $Z=3$ (sparsest graph) is studied in detail for checking the effect of the sphere discretizations when approaching the  $T=0$ limit, whereas higher values of $Z$ are studied mainly to compare with the result in the FC limit ($Z=\infty$).

\mycolor{We point out that, even though in the main text we focus on random regular graphs and fields distributed uniformly on the sphere, the techniques presented are very general and can be applied with minimal adjustments to different topologies of the network (such as Erdős–Rényi graphs) or distributions of the fields (e.g. fields whose components are distributed as Gaussians of zero mean and variance $H^2$). The changes to the methods of Sec.~\ref{sec:methods} needed to study these cases are described in App.~\ref{app:diffdegreesandfields}.}

\section{Methods}\label{sec:methods}

\mycolor{In this section we introduce the methods used to solve the Heisenberg models on sparse random graphs defined in Sec.~\ref{sec:model}. We make use of the Belief Propagation formalism, which is based on the cavity method. The non-expert interested reader can find an introduction to the cavity method in textbooks \cite{mezard2009information}.}

\mycolor{
We remind the reader that the Belief Propagation algorithm solves the saddle point equations that can be derived by the cavity method within the Replica Symmetric (RS) ansatz. The RS ansatz is known to be valid not only on tree-like graphs but also on graphs with loops as long as correlation functions decay fast enough \cite{mezard2009information}.
So, the methods described in this Section can be applied in the whole paramagnetic phase, down to the critical line \cite{Parisi_2014}.}

\mycolor{
The dAT line separates the paramagnetic RS phase from the spin glass RSB phase. In the present work, we study the critical line via the instability of the RS paramagnetic solution \mycolorsecond{against the application of a random perturbation. 
While in principle the system could develop different types of collective behavior (e.g. a ferromagnetic or a spin glass phase),
we assume the coincidence of the dAT line and the one we find, based on previous results on the XY model and for the following reasons.}
First of all, the couplings we use induce frustration and forbid the formation of ferromagnetic long-range order. 
Moreover, in the fully connected limit, the instability of the paramagnetic phase does coincide with the continuous birth of a spin glass phase \cite{sharma_and_young} and we see no reason for the nature of the phase diagram to change with the lowering of the graph degree.
It is true that in some very particular models (e.g.\ the stochastic block model with 4 communities studied in Ref.~\cite{ricci2019typology}) the continuous transition in the fully connected limit may change to a discontinuous phase transition for small graph degrees. However, we do not expect this to happen in the models we are studying here, which are purely random and not ``planted'' \cite{zdeborova2016statistical}.
}

\subsection{The Belief Propagation equations at finite temperature}\label{sec:ft}
\subsubsection{Introducing the Belief Propagation equations}\label{sec:ft_eq}

We will start by introducing the Belief Propagation (BP) equations for a model of Heisenberg spins on a graph described by Hamiltonian \eqref{eq:Model}. Let us consider a pair of neighboring nodes $i$ and $j$. The following cavity marginals can be defined
\begin{itemize}
    \item $\nu_{i \to j} (\vec{s}_i)$, the marginal on $\vec{s}_i$ in absence of node $j$,
    \item $\hat{\nu}_{j \to i} (\vec{s}_i)$, the marginal on $\vec{s}_i$ in absence of all neighboring nodes except for $j$,
\end{itemize}
which can be interpreted as the message passed from node $i$ to node $j$ and the message received by node $i$ from node $j$, respectively.
The cavity marginal $\nu$ can be related to the $\hat{\nu}$ (and vice-versa) assuming that the graph is a tree. Indeed, under this assumption, 
\begin{equation}\label{eq:BPnu}
    \nu_{i \to j} (\vec{s}_i) = \frac{1}{\mathcal{Z}_{i \to j}} e^{\beta \vec{H}_i\cdot \vec{s}_i} \prod_{k \in  \partial i \backslash j} \hat{\nu}_{k \to i}(\vec{s}_i)
\end{equation}
and
\begin{equation}\label{eq:BPnuhat}
    \hat{\nu}_{j \to i}(\vec{s}_i) = \frac{1}{\tilde{\mathcal{Z}}_{j \to i}} \int \text{d} \mu(\vec{s}_j) \; e^{ \beta J_{ij}\vec{s}_i\cdot \vec{s}_j} \nu_{j \to i}(\vec{s}_j),
\end{equation}
where $\partial i$ is the set of neighbors of vertex $i$,  the integral is over the uniform measure \mycolor{$\mu$} on the sphere of radius $\sqrt{m} = \sqrt{3}$ and $\beta = 1/T$ (the Boltzmann constant is set to 1) is the inverse temperature. Plugging \eqref{eq:BPnu} into \eqref{eq:BPnuhat} yields the self consistency equations for the $\hat{\nu}$:
\begin{equation}\label{eq:finMarg}
   \boxed{ \hat{\nu}_{j \to i}(\vec{s}_i) = \frac{1}{\hat{\mathcal{Z}}_{j \to i}} \int \text{d} \mu(\vec{s}_j) \; e^{\beta J_{ij}\vec{s}_i\cdot \vec{s}_j} e^{\beta  \vec{H}_j\cdot \vec{s}_j} \prod_{k \in  \partial j \backslash i} \hat{\nu}_{k \to j}(\vec{s}_j)}
\end{equation}
Estimates of marginals for a single variable or a pair of neighbors can be written in term of cavity marginals
\begin{equation}
b_i(\vec{s}_i) = \frac{e^{\beta \vec{H}_i\cdot \vec{s}_i}}{\mathcal{Z}_i} \prod_{j \in \partial i} \hat{\nu}_{j \to i} (\vec{s}_i) \qquad
b_{ij}(\vec{s}_i,\vec{s}_j) = \frac{e^{   \beta J_{ij}\vec{s}_i\cdot \vec{s}_j}}{\mathcal{Z}_{ij}} \nu_{i \to j} (\vec{s}_i) \nu_{j \to i} (\vec{s}_j),
\end{equation}
All $\mathcal{Z}$ in equations above are normalization factors.

The stability of the BP solution can be checked through the addition of small perturbations $\delta \nu_{i \to j}$, $\delta \hat{\nu}_{j \to i}$. Then, \eqref{eq:finMarg} becomes, up to the first order in perturbations,
\begin{equation}\label{eq:finPert}
   \boxed{\begin{aligned}\delta \hat{\nu}_{j \to i}(\vec{s}_i) = \frac{1}{\hat{\mathcal{Z}}_{j \to i}}\Big [\int \text{d} \mu(\vec{s}_j) \; e^{\beta J_{ij}\vec{s}_i\cdot \vec{s}_j} e^{\beta  \vec{H}_j\cdot \vec{s}_j} \sum_{k \in  \partial j \backslash i} \delta \hat{\nu}_{k \to j}(\vec{s}_j) \prod_{\ell \in  \partial j \backslash (i,k)} \hat{\nu}_{\ell \to j}(\vec{s}_j)  - \delta \hat{\mathcal{Z}}_{j \to i}\hat{\nu}_{j \to i}(\vec{s}_i) \Big ]\end{aligned} }
\end{equation}
where now
\begin{equation}
    \delta \hat{\mathcal{Z}}_{ \mycolor{j \to i} } \equiv \int \text{d} \mu(\vec{s}_i) \int \text{d} \mu(\vec{s}_j) \; e^{\beta J_{ij}\vec{s}_i\cdot \vec{s}_j} e^{\beta  \vec{H}_{\mycolor{j}}\cdot \vec{s}_{\mycolor{j}}} \sum_{k \in  \partial \mycolor{j \backslash i}} \delta \hat{\nu}_{k \to \mycolor{j}}(\vec{s}_{\mycolor{j}} ) \prod_{\ell \in  \partial \mycolor{j} \backslash (\mycolor{i},k)} \hat{\nu}_{\ell \to \mycolor{j}}(\vec{s}_{\mycolor{j}}).
\end{equation}

\subsubsection{Solving the Belief Propagation equations numerically}\label{sec:ft_alg}

The set of equations \eqref{eq:finMarg} can be solved numerically by starting  from a random set of marginals $\hat{\nu}^{(0)}_{j \to i}$ and setting up the recursive equation
\begin{equation}\label{eq:recursive}
    \hat{\nu}^{(t+1)}_{j \to i}(\vec{s}_i) = \frac{1}{\hat{\mathcal{Z}}^{\mycolor{(t)}}_{j \to i}} \int \text{d} \mu(\vec{s}_j) \; e^{ \beta J_{ij}\vec{s}_i\cdot \vec{s}_j} e^{\beta  \vec{H}_j\cdot \vec{s}_j} \prod_{k \in  \partial j \backslash i} \hat{\nu}^{(t)}_{k \to j}(\vec{s}_j),
\end{equation}
where $t$ denotes the step of the recursion. In this way, the fixed point (if unique) can be found simply by iterating for an equilibration time $t_\text{eq}$ large enough. The case of multiple fixed points to the BP equations is outside the scope of the present work which focuses on the critical line.

To follow this path, we introduce a discretization of the sphere made of a finite number $N_p$ of points. \mycolor{Then, each marginal is encoded in $N_p$ normalized reals, each associated to one of the $N_p$ points used to discretize the sphere. Each real corresponds to the value of the cavity marginal at the corresponding point. The $N_p$ points are chosen once at the beginning and are the same for all the marginals.} The problem of choosing such $N_p$ points is discussed at length in Ref.~\cite{delebono2024uniform}. There, we illustrate a possible way of measuring how uniform a distribution of points on the sphere is. Indeed, we look at the standard deviation of the distribution of distances between first neighbors, a feature directly related to the local ordering of the points on the sphere. After testing a series of algorithms, we find those performing better. These are the ones we will use in the paper to solve the BP equations. In particular, we mainly adopt the Lattice Point method of \cite{PhysRevLett.78.2681}, which consists of placing lattices of points on the faces of an icosahedron and then projecting them on the sphere. The points are then optimized by minimizing the power-law interaction potential, $V(r) \propto r^{-\alpha}$, between any pair of points. Again, the reader should refer to Ref.~\cite{delebono2024uniform} for more details.

\mycolor{While the Belief Propagation equations must be solved on a given graph to find the corresponding fixed point marginals and correlations, if one is interested only in the statistical properties of those marginals and correlations then the Population Dynamics (PD) approach can be used \cite{mezard2009information}. In the PD algorithm, one loses reference to a specific graph and assumes to run the BP equations on a virtually infinite graph. In practice, the PD approach can predict the probability distribution of cavity messages in the infinite size limit $\mathds{P}[\hat{\nu}_{j \to i}]$, which solves the following equation \cite{mezard2009information}}
\begin{equation}
    \mathds{P}[\hat{\nu}_{j \to i}] = \mathbb{E}_{\mathcal{G},J, \vec{H}} \int \prod_{k = 1}^{\mycolor{d_j}} (\mathcal{D} \hat{\nu}_{k \to j} \mathds{P}[\hat{\nu}_{k \to j}])\delta(\hat{\nu}_{j \to i}- \mathcal{F}_{j \to i} [\{\hat{\nu}_{k \to j}\},J_{ij}, \vec{H}_j]),
\end{equation}
where the expectation is taken over the ensemble of random regular graphs $\mathcal{G}=(V,E)$ and over the disorder (couplings $J_{ij}$ and the fields $\vec{H}_i$). The integral is computed over the set of marginals with the constrain that they must satisfy \eqref{eq:finMarg}. Here, $d_j$ is the \mycolor{excess} degree \mycolor{(the number of edges connected to a node, excluding the edge through which the node was reached)} of node $j$ (a constant for RRG) and the $\delta$ assures self-consistency according to the functions $\mathcal{F}_{j \to i}$ governing the evolution of the marginals:
\begin{equation}
    \mathcal{F}_{j \to i}[\{\hat{\nu}_{k \to j}\},J_{ij},\vec{H}_j](\vec{s}_i) \equiv \frac{1}{\hat{\mathcal{Z}}_{j \to i}} \int \text{d} \mu(\vec{s}_j) \; e^{\beta J_{ij}\vec{s}_i\cdot \vec{s}_j} e^{\beta  \vec{H}_j\cdot \vec{s}_j} \prod_{k \in  \partial j \backslash i} \hat{\nu}_{k \to j}(\vec{s}_j) 
\end{equation}
or, for a discretization of the sphere,
\begin{equation}\label{eq:evolution}
    \mathcal{F}_{j \to i}[\{\hat{\nu}_{k \to j}\},J_{ij}, \vec{H}_j](\vec{s}_i) = \frac{1}{\hat{\mathcal{Z}}_{j \to i}} \sum_{n = 1}^{N_p} \; e^{\beta J_{ij}\vec{s}_i\cdot \vec{s}_n} e^{\beta \vec{H}_j\cdot \vec{s}_n} \prod_{k \in  \partial j \backslash i} \hat{\nu}_{k \to j}(\vec{s}_n) .
\end{equation} 
Here, $n$ is an index between 1 and $N_p$ that indicates which of the points on the grid is taken into account.
Computationally, this approach is implemented by storing a population of $\mathcal{N}_\text{pop}$ marginals and then refreshing it at every step using \eqref{eq:evolution} until convergence. The procedure is described in the pseudocode in Alg.~\ref{alg:PDmarginals}.

\begin{algorithm}[thb]
\caption{Population Dynamics algorithm for marginals}\label{alg:PDmarginals}

\begin{algorithmic}[1]
  \State \mycolorsecond{Initialize $\hat{\nu}_n, \; \; n = 1, \, \dots \, , \mathcal{N}_\text{pop}$, randomly}
  \State \mycolorsecond{Set $\hat{\nu}^{(0)}_n \gets \hat{\nu}_n, \; \; n = 1, \, \dots \, , \mathcal{N}_\text{pop}$}
\Comment{\mycolorsecond{Population at time $0$}}
\For{$t =1, \, \dots \, , t_{\text{eq}}$}
\Comment{$t_{\text{eq}}$} time to reach equilibrium
  \For{$n = 1, \, \dots \, ,\mathcal{N}_\text{pop}$}
    \State Extract the field $\Vec{H}_n$ \mycolor{uniformly over the sphere}   
    \mycolor{\State Draw $Z - 1$ integers $k_1, \, \dots \, ,k_{Z-1}$} uniformly in the range $[1, \mathcal{N}_\text{pop}]$
    \State Draw a coupling \mycolor{$\{J_{n}\}$} from the coupling distribution
    \Comment{Skip if \mycolor{$J_{n} =\text{const}$}}
    \State \mycolorsecond{$\hat{\nu}_n \gets \mathcal{F}_{n}[\{\hat{\nu}_{k_i}\}_{i = 1}^{Z-1},J_{n}, \vec{H}_n]$}
  \EndFor
    \State \mycolorsecond{Set $\hat{\nu}^{(t)}_n \gets \hat{\nu}_n, \; \; n = 1, \, \dots \, , \mathcal{N}_\text{pop}$}
    \Comment{\mycolorsecond{Population at time $t$}}
\EndFor
\end{algorithmic}    
\end{algorithm}

In order to check the stability of a solution one considers a set of randomly-initialized perturbations $\delta \hat{\nu}$ and then evolves them using a discretized, recursive version of \eqref{eq:finPert}. The function governing the evolution of perturbations then becomes
\begin{equation}\label{eq:evolution_pert}
\begin{gathered}
    \hat{\mathcal{F}}_{j \to i}[\{\hat{\nu}_{k \to j}\}, \{\delta \hat{\nu}_{k \to j}\},J_{ij}, \Vec{H}_j](\vec{s}_i)  \equiv \\ \equiv  \frac{1}{\hat{\mathcal{Z}}_{j \to i}}\left [\sum_{n = 1}^{N_p} \; e^{\beta J_{ij}\vec{s}_i\cdot \vec{s}_n} e^{\beta  \vec{H}_j\cdot \vec{s}_n} \sum_{k \in  \partial j \backslash i} \delta \hat{\nu}_{k \to j}(\vec{s}_n) \prod_{\ell \in  \partial j \backslash (i,k)}  \hat{\nu}_{\ell \to j}(\vec{s}_n)  - \delta \hat{\mathcal{Z}}_{j \to i}\hat{\nu}_{j \to i}(\vec{s}_i) \right]
\end{gathered}
\end{equation}
Stability can then be checked by measuring whether the discretized $L_2$ norm of the perturbations
\begin{equation}\label{eq:pert_norm}
    \Delta(t) \equiv \frac{1}{\mathcal{N_\text{pop}}}\sum_{n = 1}^{\mathcal{N_\text{pop}}} \sqrt{\frac{1}{N_p}\sum_{\{ \vec{s}_i \}} |\delta \hat{\nu}^{(t)}_{n, \, j \to i}(\vec{s}_i) |^2},
\end{equation}
grows or decays in time. Indeed, the evolution of the norm for large times is governed by the Lyapunov factor $\lambda$ as in
\begin{equation}\label{eq:Lyapunov}
    \Delta (t) \sim \lambda^t, \quad t \gg 1.
\end{equation}
so that $\lambda = 1$ signals the onset of the RSB phase.

Since the absolute norm of the perturbations does not actually matter, perturbations can be rescaled at every step so that, before applying the evolution rule \eqref{eq:evolution_pert}, $\Delta = 1$. In this way, the value of $\Delta$ after application of $\hat{\mathcal{F}}$ gives directly the measured $\lambda$ at that step, as described in the pseudocode in Alg.~\ref{alg:linearization}.

\begin{algorithm}[thb]
\caption{Linearized BP algorithm}\label{alg:linearization}
\begin{algorithmic}[1]
  \State \mycolorsecond{Initialize $\hat{\nu}_n, \; \; n = 1, \, \dots \, , \mathcal{N}_\text{pop}$, using Alg. \ref{alg:PDmarginals}}
  \State \mycolorsecond{Initialize $\delta \hat{\nu}_n, \; \; n = 1, \, \dots \, , \mathcal{N}_\text{pop}$, randomly}
  \State \mycolorsecond{Set $(\hat{\nu}^{(0)}_n, \delta \hat{\nu}^{(0)}_n) \gets (\hat{\nu}_n, \delta \hat{\nu}_n), \; \; n = 1, \, \dots \, , \mathcal{N}_\text{pop}$}
  \Comment{\mycolorsecond{Populations at time $0$}}

\For{$t =1, \, \dots \, , t_{\text{tot}}$}
  \For{$n = 1, \, \dots \, , \mathcal{N}_\text{pop}$}
      \State Extract the field $\Vec{H}_n$ \mycolor{uniformly over the sphere}
    \mycolor{\State Draw $Z - 1$ integers $k_1, \, \dots \, ,k_{Z-1}$} uniformly in the range $[1, \mathcal{N}_\text{pop}]$
    \State Draw a coupling \mycolor{$\{J_{n}\}$} from the coupling distribution
    \Comment{Skip if \mycolor{$J_{n} =\text{const}$}}
    \State \mycolorsecond{$\hat{\nu}_n \gets \mathcal{F}_{n}[\{\hat{\nu}_{k_i}\}_{i = 1}^{Z-1},J_{n}, \vec{H}_n]$}
    \Comment{Refresh  the population}
     \State \mycolorsecond{$\delta \hat{\nu}_n \gets \hat{\mathcal{F}}_{n}[\{\hat{\nu}_{k_i}\}_{i = 1}^{Z-1}, \{\delta \hat{\nu}_{k}\},J_{n}, \Vec{H}_n]$} 
  \EndFor
  \State \mycolorsecond{Set $(\hat{\nu}^{(t)}_n, \delta \hat{\nu}^{(t)}_n) \gets (\hat{\nu}_n, \delta \hat{\nu}_n), \; \; n = 1, \, \dots \, , \mathcal{N}_\text{pop}$}
  \Comment{\mycolorsecond{Populations at time $t$}}
    \State Compute $\Delta (t)$
    \State Compute $\log \lambda(t) \gets \log \Delta (t)$
    \State \mycolorsecond{Set $\delta \hat{\nu}_n \gets \delta \hat{\nu}^{(t)}_{n} / \Delta(t), \; n = 1, \, \dots \, , \mathcal{N}_\text{pop}$}
\EndFor
\State Average $\log \lambda(t)$ to get $\log \lambda$
\end{algorithmic}
\end{algorithm}

While, in principle, $\Delta(t)$ could be measured more easily by considering two different but close populations of marginals, in practice this approach leads to saturation problems which on the contrary do not affect the linearized algorithm. Details are given in App.~\ref{app:saturation}.

\subsection{Belief Propagation at zero temperature: the max-sum equations}\label{sec:zt}

\subsubsection{Analytical expressions for the zero-temperature limit: the max-sum equations}\label{sec:zt_eq}

When moving to the $T \to 0$ limit, one needs to rewrite \eqref{eq:finMarg} and \eqref{eq:finPert} to avoid divergences. In order to do so, let us consider the new functions $h_{i \to j}$ and $u_{i \to j}$ such that
\begin{equation}
    \nu_{i \to j} = e^{\beta h_{i \to j}},
\end{equation}
\begin{equation}
    \hat{\nu}_{i \to j} = e^{\beta u_{i \to j}},
\end{equation}
and then proceed in the same way with the BP equations at finite temperature. Notice that $h_{i \to j}$ and $u_{i \to j}$ are, fundamentally, large deviation functions, and they have to be negative semi-definite (with a maximum equal to zero) in order for $\nu_{i \to j}$ and $\hat{\nu}_{i \to j}$ to be well-defined quantities. 

Since now $\beta \to \infty$, every time an integral has to be computed, it is sufficient to perform a saddle point evaluation. After normalization one gets:
\begin{equation}
    h_{i \to j}(\vec{s}_i) = \vec{H}_i \cdot \vec{s}_i + \sum_{k \in \partial i \backslash j} u_{k \to i}(\vec{s}_i) - \max_{\vec{s}_i} [\vec{H}_i \cdot \vec{s}_i + \sum_{k \in \partial i \backslash j} u_{k \to i}(\vec{s}_i) ]
\end{equation}
\begin{equation}
    u_{i \to j}(\vec{s}_j) = \max_{\vec{s}_i} [J_{ij}\vec{s}_i \cdot \vec{s}_j + h_{i \to j}(\vec{s}_i)] - \max_{\vec{s}_j} \{ \max_{\vec{s}_i} [J_{ij}\vec{s}_i \cdot \vec{s}_j + h_{i \to j}(\vec{s}_i)] \}.
\end{equation}
As predicted, normalization makes the $h$ and $u$ functions negative semi-definite. Putting together the previous equations one gets the \textit{max-sum equations}:
\begin{equation}\label{eq:maxsum_u}
    \boxed{ u_{i \to j}(\vec{s}_j) =  \max_{\vec{s}_i} [J_{ij}\vec{s}_i \cdot \vec{s}_j + \vec{H}_i \cdot \vec{s}_i + \sum_{k \in \partial i \backslash j} u_{k \to i}(\vec{s}_i)] + \text{normalization}},
\end{equation}
where the additional normalization constant is given by $-\max_{\vec{s}_j} u_{i \to j}$.
Adding a small perturbation \mycolor{to~\eqref{eq:maxsum_u} (or, equivalently, taking the zero-temperature limit of~\eqref{eq:finPert}, since the two operations commute, as shown in App.~\ref{app:commuting})} leads to 
\begin{equation}\label{eq:maxsum_deltau}
    \boxed{ \delta u_{i \to j}(\vec{s}_j) = \sum_{k \in \partial i \backslash j} \delta u_{k \to i}(\vec{s}_i^{\, *}(\vec{s}_j)) + \text{normalization}}
\end{equation}
The additive normalization constant makes the perturbation equal to zero in correspondence of the largest (i.e. the null) marginal and
\begin{equation}\label{eq:argmax}
    \vec{s}_i^{\, *}(s_j) \equiv \text{argmax}_{\vec{s}_i} [J_{ij}\vec{s}_i \cdot \vec{s}_j + \vec{H}_i \cdot \vec{s}_i + \sum_{k \in \partial i \backslash j} u_{k \to i}(\vec{s}_i)].
\end{equation}

In the end, the functions governing the evolution are
\begin{equation}
    \mathcal{F}_{j \to i}[\{\hat{\nu}_{k \to j}\},J_{ij}, \vec{H}_j](\vec{s}_i) \equiv  \max_{\vec{s}_i} [J_{ij}\vec{s}_i \cdot \vec{s}_j + \vec{H}_i \cdot \vec{s}_i + \sum_{k \in \partial i \backslash j} u_{k \to i}(\vec{s}_i)] + \text{normalization}
\end{equation}
and
\begin{equation}
    \hat{\mathcal{F}}_{j \to i}[\{\hat{\nu}_{k \to j}\}, \{\delta \hat{\nu}_{k \to j}\},J_{ij}, \vec{H}_j](\vec{s}_i) \equiv  \sum_{k \in \partial i \backslash j} \delta u_{k \to i}(\vec{s}_i^{\, *}(\vec{s}_j))+ \text{normalization}.
\end{equation}

\subsubsection{Numerical implementation of the max-sum algorithm}\label{sec:BPT0_alg}\label{sec:zt_alg}

\mycolor{Any} discretization of \eqref{eq:maxsum_u} and \eqref{eq:maxsum_deltau} \mycolor{may in principle lead to inaccurate estimates for the marginals, as it happens in the XY model \cite{lupo2017approximating, lupo2019random} (and we find that this is indeed the case,} see Figs.~\ref{fig:logl_diffr} and ~\ref{fig:h_vs_r}). Therefore, it is necessary to apply an interpolation procedure, as successfully done in the XY case \cite{lupo2017approximating, lupo2019random}. 

While for XY variables the interpolation was easy because any one-dimensional function can be well approximated by a simple parabola close to its maximum, when dealing with Heisenberg spins the interpolation process is more complicated because it has to be carried out in a two-dimensional space \mycolor{(see App.~\ref{app:additional_complications}), as explained in~\ref{app:max_with_interpolation}}. 
The algorithm is summarized by the pseudocode in Alg.~\ref{alg:maxsum}.

\begin{algorithm}[thb]
\caption{Max-sum algorithm}\label{alg:maxsum}
\begin{algorithmic}[1]
\State \mycolorsecond{Initialize $\hat{\nu}_n, \; \; n = 1, \, \dots \, , \mathcal{N}_\text{pop}$, randomly}
\State \mycolorsecond{Set $\hat{\nu}^{(0)}_n \gets \hat{\nu}_n, \; \; n = 1, \, \dots \, , \mathcal{N}_\text{pop}$}
  \Comment{\mycolorsecond{Population at time $0$}}

\For{$t =1, \, \dots \, , t_{\text{eq}}$} \Comment{Find the fixed point for the marginals}
  \For{$n = 1, \, \dots \, ,\mathcal{N}_\text{pop}$}
      \State Extract the field $\Vec{H}_n$ \mycolor{uniformly over the sphere}
    \mycolor{\State Draw $Z - 1$ integers $k_1, \, \dots \, ,k_{Z-1}$} uniformly in the range $[1, \mathcal{N}_\text{pop}]$
    \State Draw a coupling \mycolor{$\{J_{n}\}$} from the coupling distribution
    \Comment{Skip if \mycolor{$J_{n} =\text{const}$}}
     \mycolorsecond{\State $\hat{\nu}_n \gets \mathcal{F}_{n}[\{\hat{\nu}_{k_i}\}_{i = 1}^{Z-1},J_{n}, \vec{H}_n]$}
  \EndFor
  \State \mycolorsecond{Set $\hat{\nu}^{(t)}_n \gets \hat{\nu}_n, \; \; n = 1, \, \dots \, , \mathcal{N}_\text{pop}$}
  \Comment{\mycolorsecond{Population at time $t$}}
  
\EndFor
  \State \mycolorsecond{Initialize $\delta \hat{\nu}_n, \; \; n = 1, \, \dots \, , \mathcal{N}_\text{pop}$, randomly}
  \State \mycolorsecond{Set $\delta \hat{\nu}^{(0)}_n \gets \delta \hat{\nu}_n, \; \; n = 1, \, \dots \, , \mathcal{N}_\text{pop}$}
  \Comment{\mycolorsecond{Population at time $0$}}
  
\For{$t =1, \, \dots \, , t_{\text{tot}}$}
  \For{$n = 1, \, \dots \, ,\mathcal{N}_\text{pop}$}
      \State Extract the field $\Vec{H}_n$ \mycolor{uniformly over the sphere}
    \mycolor{\State Draw $Z - 1$ integers $k_1, \, \dots \, ,k_{Z-1}$} uniformly in the range $[1, \mathcal{N}_\text{pop}]$
    \State Draw a coupling \mycolor{$\{J_{n}\}$} from the coupling distribution
    \Comment{Skip if \mycolor{$J_{n} =\text{const}$}}
    \State Interpolate using marginals
    \State \mycolorsecond{$\hat{\nu}_n \gets \mathcal{F}_{n}[\{\hat{\nu}_{k_i}\}_{i = 1}^{Z-1},J_{n}, \vec{H}_n]$}
    \Comment{Using interpolated maximum}
     \State \mycolorsecond{$\delta \hat{\nu}_n \gets \hat{\mathcal{F}}_{n}[\{\hat{\nu}_{k_i}\}_{i = 1}^{Z-1}, \{\delta \hat{\nu}_{k_i}\}_{i = 1}^{Z-1},J_{n}, \vec{H}_n]$} \Comment{Using interpolated argmax}
     \State Normalize \mycolorsecond{$\hat{\nu}_{n}$} using interpolation
    \State Normalize \mycolorsecond{$\delta \hat{\nu}_{n}$} using interpolation
  \EndFor
  \State \mycolorsecond{Set $(\hat{\nu}^{(t)}_n, \delta \hat{\nu}^{(t)}_n) \gets (\hat{\nu}_n,\delta \hat{\nu}_n), \; \; n = 1, \, \dots \, , \mathcal{N}_\text{pop}$}
  \Comment{\mycolorsecond{Populations at time $t$}}
    \State Compute $\Delta (t)$
    \State Compute $\log \lambda(t) \gets \log \Delta (t)$
    \State \mycolorsecond{$\delta \hat{\nu}_n \gets \delta \hat{\nu}^{(t)}_{n} / \Delta(t)$}
\EndFor
\State Average $\log \lambda(t)$ to get $\log \lambda$
\State Obtain $H_c$ via a linear fit and checking when $\log \lambda = 0$
\end{algorithmic}
\end{algorithm}

\subsection{The large-connectivity limit}\label{sec:large}

\subsubsection{The large connectivity expansion}\label{sec:large_eq}

To check that our generic computation recovers in the limit $Z\to\infty$ the results obtained for fully connected (FC) models we can solve the BP equations via the vectorial ansatz introduced in \cite{javanmard2016phase}. Indeed,  in this limit, couplings must scale like $J_{ij}=\mathcal{O}(1/\sqrt{Z})$, and thus we can expand the BP equation \mycolor{\eqref{eq:BPnuhat}} obtaining
\bea\label{eq:largeZ_nuhat}
\hat\nu_{i\to j}(\bs_j) \propto 1 + \beta J_{ij} \int \dd \mu (\bs_i) \nu_{i\to j}(\bs_i)  \bs_i \cdot \bs_j \propto \exp(\beta J_{ij} \vec{m}_{i\to j} \cdot \vec{s}_j),
\eea
where we have introduced the \textit{cavity magnetizations}
\begin{equation}\label{eq:largeZ_m}
    \bmm_{i\to j} \equiv \int \dd \mu (\bs_i) \nu_{i\to j}(\bs_i) \bs_i.
\end{equation}
The other cavity marginal \eqref{eq:BPnu} can be written in exponential form as well
\bea\label{eq:largeZ_nu}
\nu_{i\to j}(\bs_i) \propto \exp(\beta \bh_{i\to j} \cdot \bs_i).
\eea
Then one can write self-consistent equations to compute the parameters $\vec{m}_{i\to j}$ and $\bh_{i\to j}$ and, from them, calculate the location of the dAT line (see App.~\ref{app:vectorial_ansatz} for details). The result in the $Z=\infty$ limit is equivalent to the computation by Sharma and Young \cite{sharma_and_young}.

The advantage of deriving this limit from the BP equations is that we can also access finite $Z$ corrections by adding additional terms to the expansion in $\beta J_{ij}$. One can then write a set of closed equations for $\nu_{i \to j}$ and $\hat{\nu}_{i \to j}$ such that cavity marginals are in the form of an exponential of a polynomial of a given degree of the spins (see App.~\ref{app:expansion} for details). The first correction to the vectorial ansatz is of order $\mathcal{O}(1/Z)$ and requires to go up to fourth powers of $\beta J_{ij}$ (App.~\ref{app:expansion}). Once one has computed all the combinatorial terms at a given power of $1/Z$, one can set up a numerical procedure to find the polynomial coefficients via the Population Dynamics algorithm. Although averages such as \eqref{eq:largeZ_m} cannot be computed analytically for finite $Z$ (at variance to the leading term), these can be computed numerically using the usual sphere discretization.

\subsubsection{Asymmetric Stability Checking (ASC) algorithm for the large connectivity expansion}\label{sec:large_alg}

Finding the dAT line via the linearized procedure requires using the linearized BP equations, which in principle can be messy to compute and, in particular, are order-dependent, so adding additional terms to the $1/Z$ expansion requires a modification of the equations.

We therefore introduce a novel algorithm that avoids the problem of computing the linearized equations. The method is based on the well-known technique \cite{pagnani2003near} of evolving two close-by populations\mycolor{, $\hat{\nu}_{j \to i}$ and $\hat{\mu}_{j \to i}$}, and checking whether, during the evolution, they tend to come closer or farther apart\mycolor{, as measured by the distance-between-populations version of~\eqref{eq:pert_norm}:
\begin{equation}
        \Delta(t) \equiv \frac{1}{\mathcal{N_\text{pop}}}\sum_{n = 1}^{\mathcal{N_\text{pop}}} \sqrt{\frac{1}{N_p}\sum_{\{ \vec{s}_i \}} |\hat{\nu}^{(t)}_{n, \, j \to i}(\vec{s}_i) - \hat{\mu}^{(t)}_{n, \, j \to i}(\vec{s}_i) |^2},
\end{equation}
}
To correctly perform this check we need to keep the two populations at the proper distance, in particular avoiding the saturation problems described in App.~\ref{app:saturation}. 

\begin{algorithm}[thb]
\caption{Asymmetric Stability Checking (ASC) algorithm}\label{alg:move}
\begin{algorithmic}[1]
  \State Initialize \mycolorsecond{$\hat{\nu}_{n}, \; n = 1, \, \dots \, , \mathcal{N}_\text{pop} $,} using Alg. \ref{alg:PDmarginals}
  \State Initialize \mycolorsecond{$\hat{\mu}_{n}, \; n = 1, \, \dots \, , \,\mathcal{N}_\text{pop},$} randomly so that $\Delta = \epsilon \ll 1$
  \State \mycolorsecond{Set $(\hat{\nu}^{(0)}_n, \hat{\mu}^{(0)}_n) \gets (\hat{\nu}_n, \hat{\mu}_n), \; \; n = 1, \, \dots \, , \mathcal{N}_\text{pop}$}
  \Comment{\mycolorsecond{Populations at time $0$}}
  \State Set $\tilde{\Delta}(0) = 1$

\For{$t =1, \, \dots \, , t_{\text{tot}}$}
  \For{$n = 1, \, \dots \, ,\mathcal{N}_\text{pop}$}
      \State Extract the field $\Vec{H}_j$ \mycolor{uniformly over the sphere}
    \mycolor{\State Draw $Z - 1$ integers $k_1, \, \dots \, ,k_{Z-1}$} uniformly in the range $[1, \mathcal{N}_\text{pop}]$
    \State Draw a coupling \mycolor{$\{J_{n}\}$} from the coupling distribution
    \Comment{Skip if \mycolor{$J_{n} =\text{const}$}}
\State \mycolorsecond{$\hat{\nu}_n \gets \mathcal{F}_{n}[\{\hat{\nu}_{k_i}\}_{i = 1}^{Z-1},J_{n}, \vec{H}_n]$}
\State \mycolorsecond{$\hat{\mu}_n \gets \mathcal{F}_{n}[\{\hat{\mu}_{k_i}\}_{i = 1}^{Z-1},J_{n}, \Vec{H}_n]$}
  \EndFor
  \State \mycolorsecond{Set $(\hat{\nu}^{(t)}_n, \hat{\mu}^{(t)}_n) \gets (\hat{\nu}_n, \hat{\mu}_n), \; \; n = 1, \, \dots \, , \mathcal{N}_\text{pop}$}
  \Comment{\mycolorsecond{Populations at time $t$}}
       \State Compute $\lambda(t) \gets \Delta (t) / \tilde{\Delta}(t-1)$
       \If{$\lambda(t)$ > 1}
       \State \mycolor{$M_n \gets (\hat{\nu}_{n} +\hat{\mu}_{n})/2$}
       \State \mycolor{$D_n \gets (\hat{\nu}_{n}-\hat{\mu}_{n})/2$}
       \State \mycolor{$\hat{\nu}_n \gets M_{n} + D_{n}/ \lambda(t)$}
       \Comment{Move the populations closer together}
        \State \mycolor{$\hat{\mu}_n \gets M_{n} - D_{n}/ \lambda(t)$}
       \Else 
       \State Set $\tilde{\Delta}(t) \gets \Delta (t)$
       \EndIf
    \EndFor
\State Average $\log \lambda (t)$ to obtain $\log \lambda$
\end{algorithmic}
\end{algorithm}

Our algorithm, which we call Asymmetric Stability Checking (ASC), works as follows.
Let us call $\{\hat{\nu}_k,\hat{\mu}_k\}$ the two populations of cavity marginals and $\lambda$ the Lyapunov factor of \eqref{eq:Lyapunov} measured in the last step. 
Our idea is to renormalize the two populations in a way such that they can not go too far apart, thus leaving the regime of linear response.
For every element of the two populations, we apply the following transformation only if $\lambda>1$
\begin{equation} \label{ASC}
    \hat{\nu}_k \leftarrow M_k + \frac{D_k}{\lambda}\;, \quad
    \hat{\mu}_k \leftarrow M_k - \frac{D_k}{\lambda}\;, \qquad
    \text{with} \qquad
    M_k \equiv \frac{\hat{\nu}_k +\hat{\mu}_k}{2}\;, \quad
    D_k \equiv \frac{\hat{\nu}_k -\hat{\mu}_k}{2}\;.
\end{equation}
The above transformation avoids any divergence in the distance between the two populations while preserving the small difference between the two which permits the measure of the largest eigenvalue of the linear operator close to the BP fixed point.

If $\lambda > 1$, the transformation in (\ref{ASC}) decreases the distance between the two populations by the same factor $\lambda$ that increased during the last step of the population dynamics, which is the goal of the procedure. The transformation is linear are keeps both marginals well normalized.

The reason why we can not apply transformation in (\ref{ASC}) when $\lambda<1$ is because the new marginals may become negative, which is clearly unacceptable.
Nonetheless when $\lambda<1$ the two populations tend to get closer and this is not a problem until the differences are larger than numerical precision (and we always work in that regime). The only needed caution is to redefine the baseline value for $\Delta(t)$.

\mycolor{We point out that the procedure described above is quite general and can also be applied to the finite-connectivity case.}

The whole procedure is described in Alg. \ref{alg:move}. 

\section{Results}\label{sec:results}

All results in this section have been obtained using a population of $\mathcal{N}_\text{pop} = 10^4$ marginals. Moreover, couplings have been taken as $J_{ij} = -\frac{1}{\sqrt{Z-1}}$. Strictly speaking, this corresponds to the antiferromagnetic case. However, since we work on RRG, we expect half of the loops to be frustrated, exactly as in the case of the random couplings. Hence, the location of the dAT line is the same, but the computational complexity is lower. 

\mycolor{We note that using antiferromagnetic couplings when using the Population Dynamics approach can, in principle, cause problems in those models that can be transformed related to a ferromagnetic one by a gauge transformation. For instance, the population can jump between states, each related to the other by a given symmetry. However, as reported in the pseudocodes of Sec. \ref{sec:methods}, we update the population in a sequential order, and not in a parallel one. This kind of update rule avoids the oscillation in the population between different states. 
}

\subsection{Finite temperature, finite connectivity results}\label{sec:res_ft}

In order to get the critical field defining the dAT line as a function of temperature, $H_c(T)$, one can simply compute the logarithm of the Lyapunov factor as a function of $H$ using Alg. \ref{alg:linearization} and then perform an interpolation (either linear or quadratic) to find the value $H_c$ at which $\log \lambda = 0$.

\begin{figure}[t] 
    \centering
    \makebox[\textwidth][c]{\includegraphics[width=0.6\textwidth]{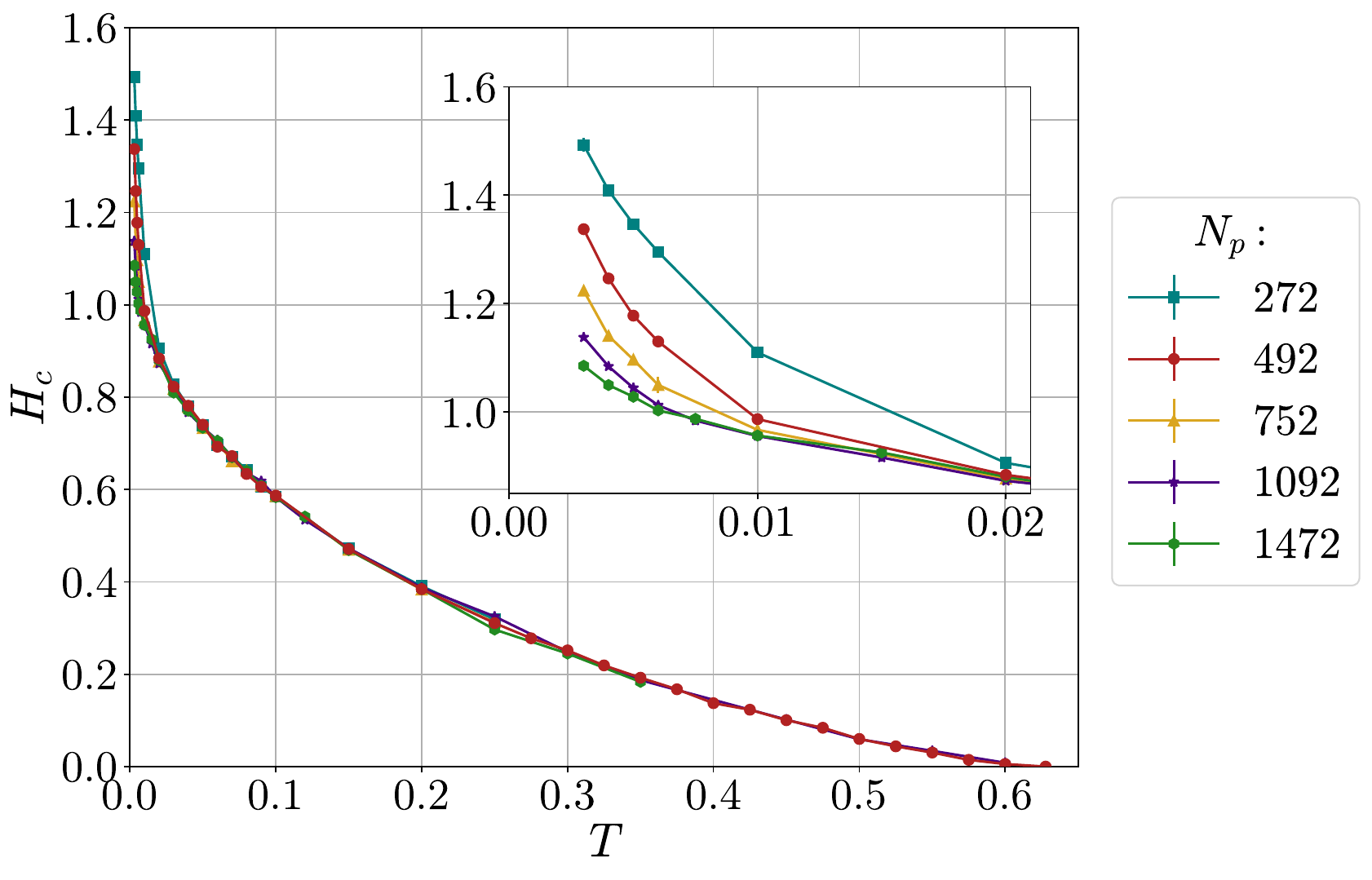}}%
    \caption[dAT line for the spin glass Heisenberg model on a RRG of degree $Z = 3$]{\textit{Main plot:} dAT line for the spin glass Heisenberg model on a RRG of degree $Z = 3$. Different sizes of the sphere grid are considered. \textit{Inset}: zoom-in of the low-$T$, high-$H$ region, in which discretization effects start to become important. Error bars have sizes comparable to data points. Lines are just a guide for the eye.}
    \label{fig:dat}
\end{figure}

We considered RRG of connectivity $Z = 3$ and different number of points $N_p$ to discretize the sphere. Results are presented in Fig. \ref{fig:dat}. The curves for different $N_p$ collapse onto each other for large enough values of the temperature. However, finite-size effects start to appear when $T$ is decreased very much. This is reasonable, since finite-grid effects are known to appear in other models, e.g. differentiating between the clock model with a finite number of states and the XY model \cite{lupo2017approximating}. Going to lower temperatures requires using more points on the sphere, and eventually changing to Alg.~\ref{alg:maxsum} to reach the $T \to 0$ limit.

To obtain the critical temperature at zero field, $T_g$, we run the BP algorithm without an external field. We can consider a rather rough discretization, namely one with $N_p = 272$, since finite-grid effects only appear at low temperatures, as previously shown. The critical temperature found from this approach is
\begin{equation}
\label{eq:zero_field_crit_temp_num_result}
    T_g = 0.627 \pm 0.001.
\end{equation}
The critical temperature $T_g$ can be obtained also analytically as the solution of the following equation
\begin{equation}
\label{eq:equation_for_Tg}
    (Z-1)\left[\coth\left(\frac{3}{\sqrt{Z-1}\,T_g}\right)-\frac{\sqrt{Z-1}\,T_g}{3}\right]^2=1,
\end{equation}
\mycolor{that comes from the results in \cite{coolen2005finitely}, as explained in detail in App.~\ref{app:_risultati_articolo}}.
We find for $Z=3$
\begin{equation}
\label{eq:zero_field_crit_temp_theor_result}
    T_g^*\,\approx\,0.626
\end{equation}
which perfectly agrees with the numerical result.
\mycolor{This agreement on the zero-field critical temperature is very satisfying given that the analytical approach of Ref.~\cite{coolen2005finitely} and our numerical computation are different. They provide the same answer for $H=0$ because both probe the entire space of possible perturbations (in the highly symmetric paramagnetic phase without field). For $H\neq 0$ only our numerical approach is viable.}

\begin{figure}[t] 
    \centering
    \makebox[\textwidth][c]{\includegraphics[width=0.5\textwidth]{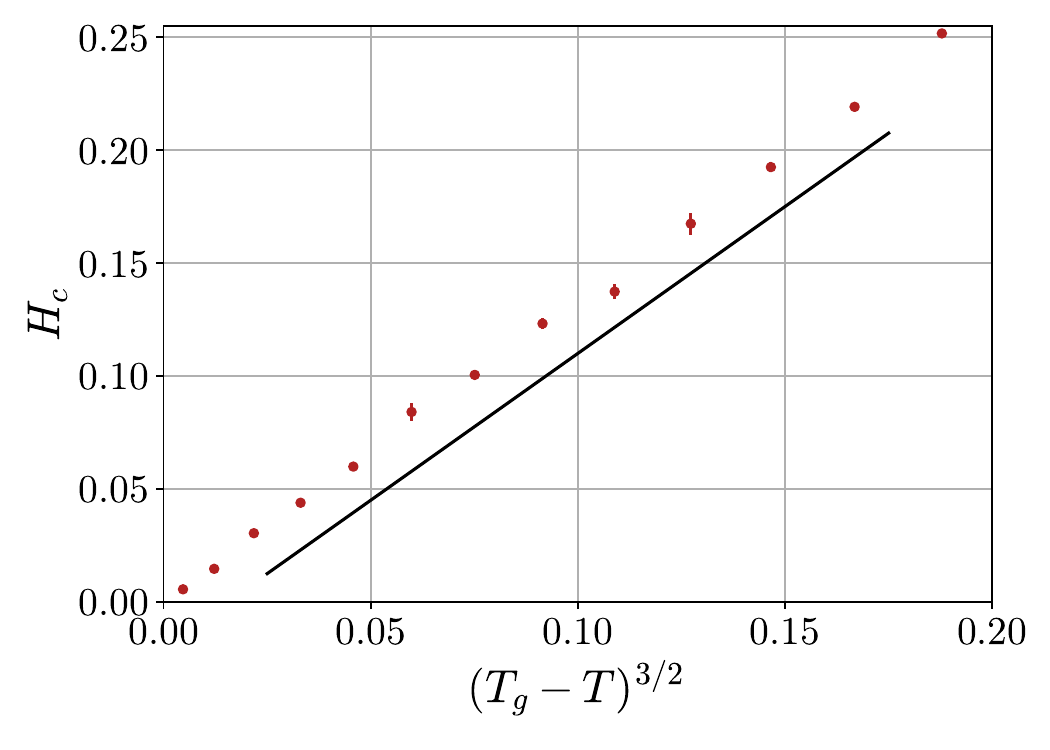}}%
    \caption[Critical field $H_c(T)$ as a function of $(T_g-T)^\frac{3}{2}$]{Critical field $H_c(T)$ as a function of $(T_g-T)^\frac{3}{2}$. The black line acts as a comparison. The behavior of the dAT line is clearly linear in this region. $N_p = 492$.}
    \label{fig:3_2exponent}
\end{figure}

When moving at temperatures $T \lesssim T_g$, one expects the behavior to be in the form $H \propto (T_g-T)^{3/2}$, as in the XY model \cite{lupoTesi}. Indeed, as shown in Fig. \ref{fig:3_2exponent}, the value of $3/2$ of the critical exponent is correct also in the case of Heisenberg spins.

\begin{figure}[t] 
    \centering
    \makebox[\textwidth][c]{\includegraphics[width=0.5\textwidth]{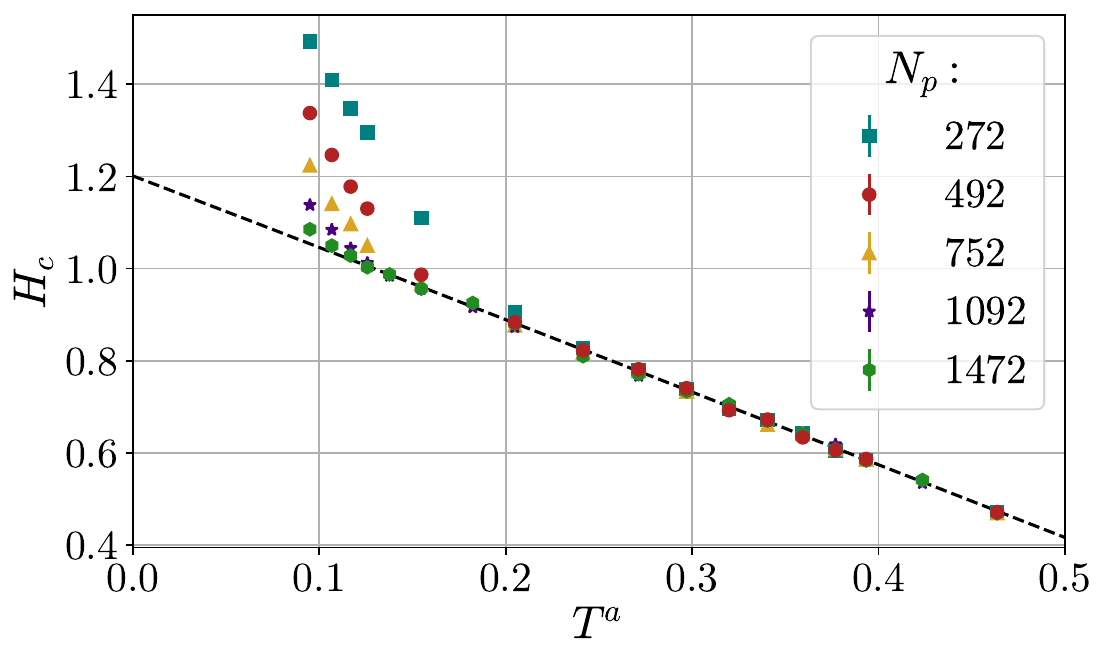}}%
    \caption[Critical field of the dAT line as a function of $T^a$]{Critical field of the dAT line as a function of $T^a$, where $a = 0.405$, together with the infinite-grid extrapolation. Error bars have sizes comparable to data points.}
    \label{fig:fit_lin}
\end{figure}

Finally, as previously stated, when going to the zero temperature limit, discretization effects start to become important. In order to compute the critical field at zero temperature, therefore, we have to resort to extrapolation from finite $T$ (Fig. \ref{fig:fit_lin}).
Details of the extrapolation are presented in App.~\ref{app:extrapolation}. The extrapolation yields
\begin{equation}
    H_c(0) = 1.20 \pm 0.02
\end{equation}
as our estimate of the zero temperature critical field.

\subsection{Zero temperature results}\label{sec:res_0T}

\begin{figure}[t] 
    \centering
    \makebox[\textwidth][c]{\includegraphics[width=0.7\textwidth]{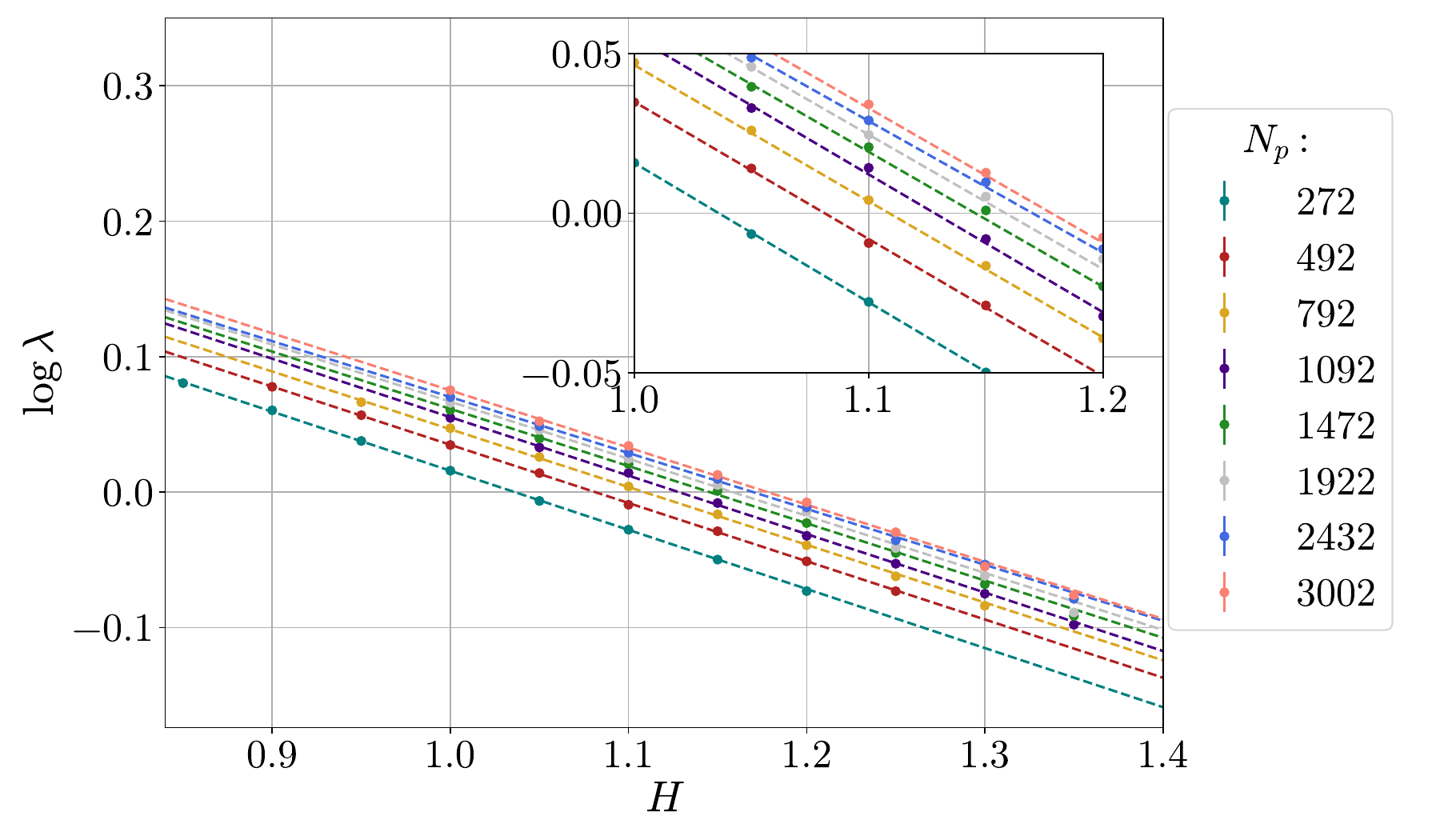}}%
    \caption[Behaviour of $\log \lambda$ vs $H$ at $T = 0$ for different values of $N_p$]{\textit{Main plot:} behaviour of $\log \lambda$ vs $H$ at $T = 0$ for different values of $N_p$ ($V$ potential with $\alpha = 1$). \textit{Inset:} zoom-in of the $\log \lambda \simeq 0$ region of the plot. Error bars are smaller than data points. Dashed lines are linear fits. For each $N_p$, the highest-$H$ point has been ignored in order to avoid non-linear behaviours. In order to reduce noise, all the results have been obtained by averaging together ten different runs of $t_\text{tot} = 500$  steps each ($t_\text{ther} = 100$ steps were excluded to take into account thermalization effects).}
    \label{fig:logl_maxsum}
\end{figure}

The critical field at zero temperature found using the extrapolation procedure can be checked against that found using the algorithm at zero temperature described in Sec. \ref{sec:BPT0_alg}.
We ran the algorithm for different values of $N_p$. In Fig. \ref{fig:logl_maxsum}, the resulting $\log \lambda$ vs $H$ plot is reported. 

\begin{figure}[t] 
    \centering
    \makebox[\textwidth][c]{\includegraphics[width=0.55\textwidth]{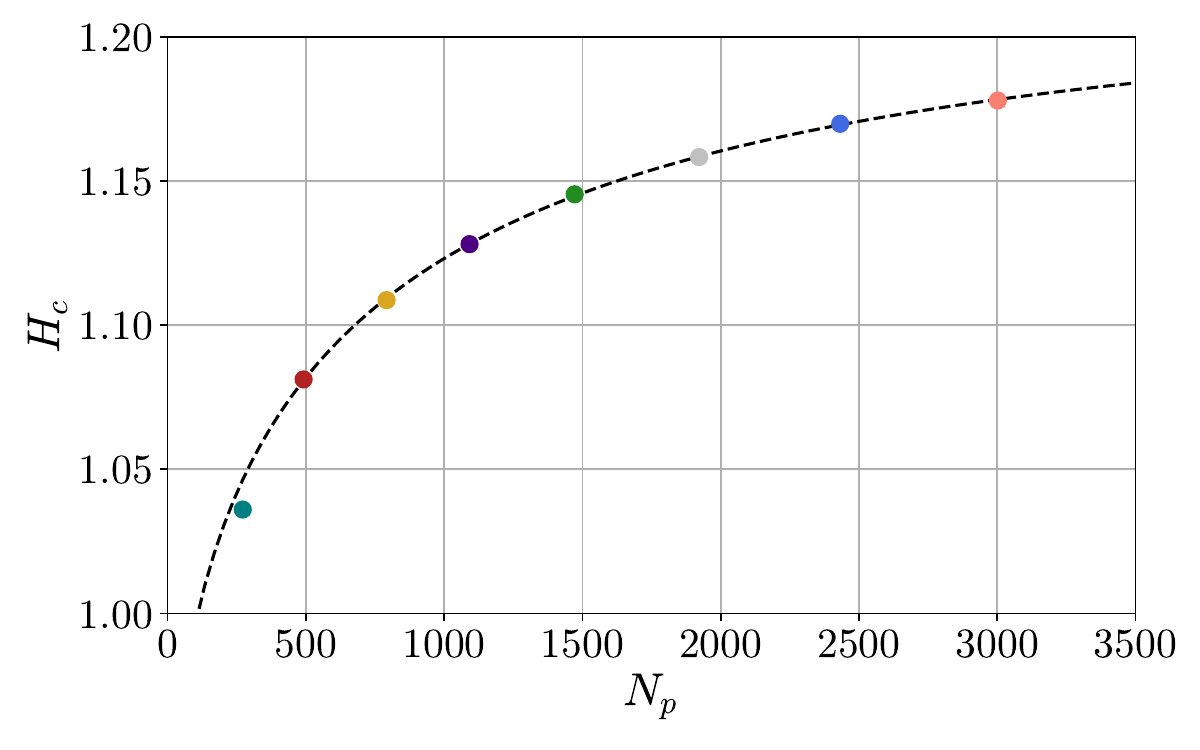}}%
    \caption[Critical field at zero temperature as a function of $N_p$]{Critical field at zero temperature, $H_c(T = 0)$, as a function of $N_p$. Colours of data points are chosen in order to match those in Fig. \ref{fig:logl_maxsum}. The dashed line was obtained via a fit to the function defined in \eqref{eq:exp_fit}, excluding $N_p = 272$. Error bars are smaller than data points.}
    \label{fig:maxsum_fitexp}
\end{figure}

\begin{figure}[t] 
    \centering
    \makebox[\textwidth][c]{\includegraphics[width=0.55\textwidth]{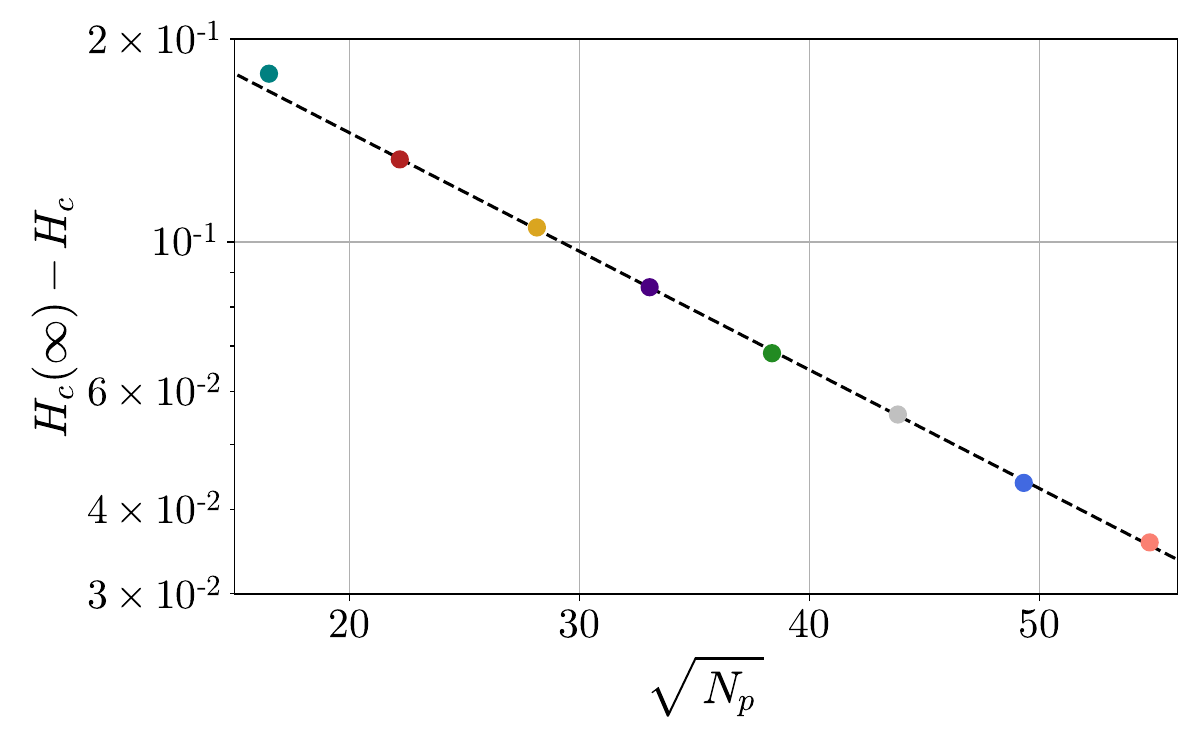}}%
    \caption[Difference between the critical field for $N_p (\infty)$ and that at finite $N_p$ as a function of $\sqrt{N_p}$]{Difference between the critical field for $N_p (\infty)$ and that at finite $N_p$ as a function of $\sqrt{N_p}$ in semi-log scale. Colors of data points are chosen to match those in Fig. \ref{fig:logl_maxsum}. The dashed line is a fit to the function defined in \eqref{eq:exp_fit}. Error bars are smaller than data points.}
    \label{fig:maxsum_deltaH}
\end{figure}

\begin{figure}[t] 
    \centering
    \makebox[\textwidth][c]{\includegraphics[width=0.55\textwidth]{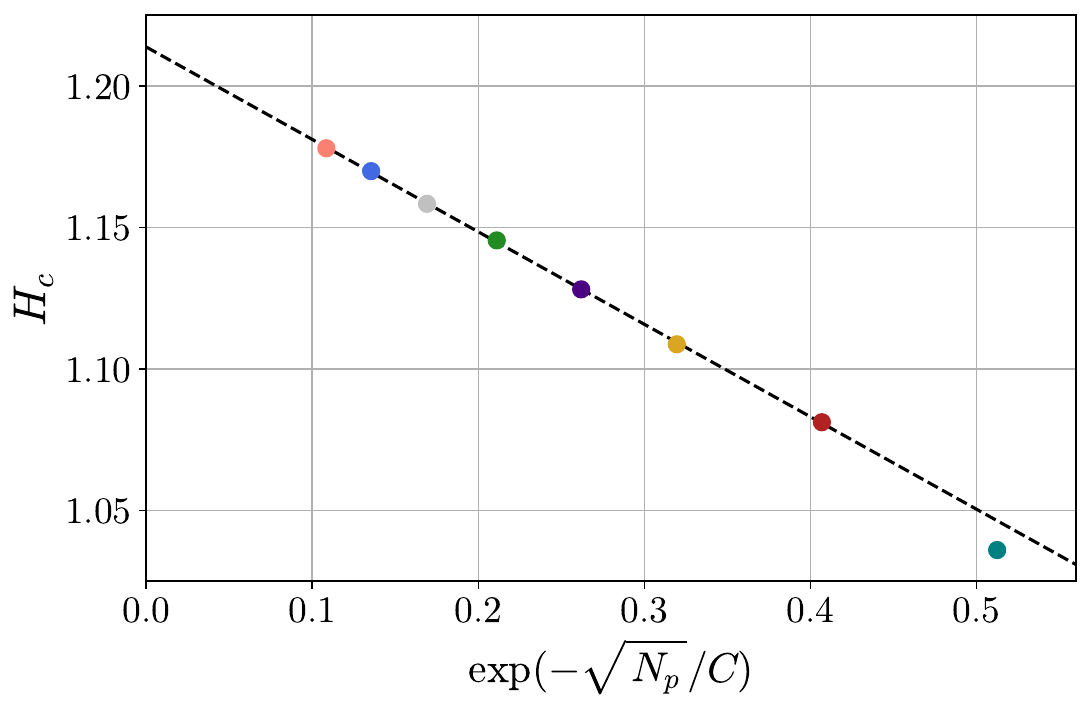}}%
    \caption[Critical field at zero temperature, $H_c(T = 0)$, as a function of $\exp(-\sqrt{N_p}/C)$]{Critical field at zero temperature, $H_c(T = 0)$, as a function of $\exp(-\sqrt{N_p}/C)$. Colors of data points are chosen to match those in Fig. \ref{fig:logl_maxsum}. The dashed line is a fit to the function defined in \eqref{eq:exp_fit}. Error bars are smaller than data points.}
    \label{fig:maxsum_extrapolation}
\end{figure}

The scaling of $H_c(T = 0)$ as a function of $N_p$ is depicted in Fig.~\ref{fig:maxsum_fitexp}. There is still a dependence on the number of points $N_p$ but the convergence to the exact value ($N_p\to\infty$) is very fast and can be very well fitted by the following law
\begin{equation}\label{eq:exp_fit}
    H_c(N_p) = H_c(\infty) +  B \exp(-\sqrt{N_p}/C)\;.
\end{equation}
The $\sqrt{N_p}$ in the argument of the exponential can be justified by noticing that on a sphere, the average distance between points scales as the inverse of the square root of the number of points, as in \eqref{eq:char_len_0}. The lowest value of $N_p$, that is $N_p = 272$, has been excluded from the fitting procedure, to avoid small-grid effects. The parameters estimates resulting from the fitting procedures are: 
\begin{equation}\label{eq:fit_parameter_0T}
    \begin{gathered}
        H_c(\infty) = 1.214 \pm 0.003 \\
B = -0.327\pm 0.005 \\
C = 24.7 \pm 0.9 \\
    \end{gathered}
\end{equation}
To check the goodness of this fit, the behavior of $H_c(\infty) - H_c(N_p)$ as a function of $\sqrt{N_p}$ in a semi-log plot is shown in Fig.~\ref{fig:maxsum_deltaH}. Moreover, the scaling of $H_c$ as a function of $\exp(-\sqrt{N_p}/C)$, together with the extrapolation to infinite $N_p$, is depicted in Fig.~\ref{fig:maxsum_extrapolation}. In both cases, the behavior is clearly linear in the range of $N_p$ considered.

Taking into account also the $N_p = 272$ point does not appear to change significantly the visual goodness of the fit, but it does increase the $\chi^2$ value. The corresponding critical field, however, is $H_c = 1.199 \pm 0.006$, not very far off from the previous estimate. Moreover, excluding additional points from the fit changes only slightly the value of $H_c(N_p = \infty)$. In the end, the field resulting from the fit \eqref{eq:fit_parameter_0T} appears to be a good estimate of the critical field at zero temperature:
\begin{equation}
    H_c(T= 0) = 1.214 \pm 0.003.
\end{equation}
A more careful analysis could be performed in order to have a better estimate of the uncertainty. However, even as it is, this value of the critical field is in complete agreement with the result $H_c(T = 0) = 1.20 \pm 0.02$ presented above.

\subsection{Large connectivity limit}\label{sec:res_large}

\begin{figure}[t]
    \centering
    \makebox[\textwidth][c]{\includegraphics[width=0.6\textwidth]{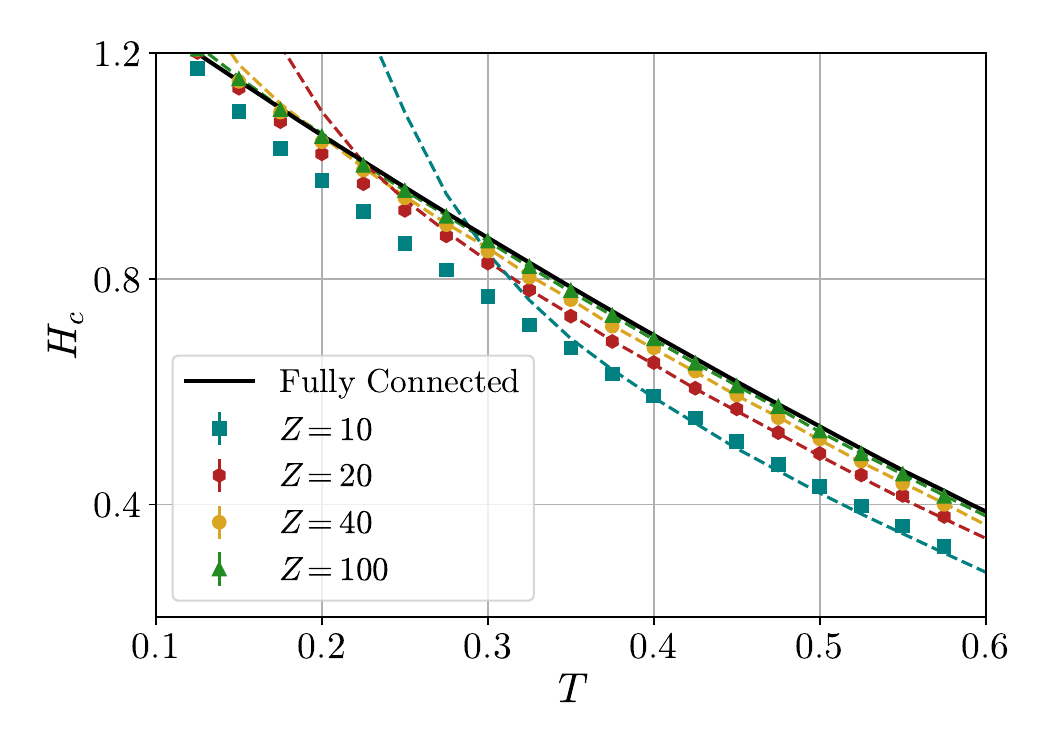}}%
    \caption[Comparison between the finite-connectivity and the FC dAT lines]{Comparison between the dAT lines for different connectivities $Z$ of the RRG (data points) and the solution for the fully connected topology (full black line). Dashed lines are the dAT lines obtained running the ASC algorithm with the quartic ansatz. Error bars are smaller than data points. Parameters: $t_\text{eq} = 30$, $t_\text{therm} = 50$, $t_\text{tot} = 100$, $N_p = 272$ ($V$ potential with $\alpha = 7$).}
    \label{fig:lines_diffZ}
\end{figure}

Finally, we move to the large connectivity limit. In Fig.~\ref{fig:lines_diffZ} we compare the dAT lines obtained using Alg.~\ref{alg:linearization} for different connectivities $Z$ (data points) to the one obtained in the FC limit (black line). \mycolor{The latter can be found via a cavity computation (see App.~\ref{app:vectorial_ansatz}, where we show how to find the general equation for the location of the line \eqref{eq:datLineEQ} and how to solve it when the external field is Gaussian or uniformly distributed)}. The agreement improves as the connectivity increases. Moreover, with dashed lines, we also report the approximated dAT lines obtained using Alg.~\ref{alg:move} with the \mycolor{quartic} ansatz for the form of the marginals (but still using the discretization). The agreement is excellent at high temperatures, thus showing that a quartic ansatz is powerful enough to describe the marginals at large connectivities. At low temperatures the disagreement increases, as can be expected by an expansion in terms of $\beta J_{ij}$. 

\section{Discussion}\label{sec:conclusions}
In this work, we studied extensively the Heisenberg spin glass model with a random field on sparse random graphs. 

Firstly, we have introduced a way to solve the Belief Propagation equations at finite temperatures and connectivities exploiting a discretization of the sphere. We have shown how perturbations can be studied via a linearized form of the Belief Propagation equations and used these results to compute the de Almeida-Thouless line in the field-temperature plane (see Fig.~\ref{fig:dat}). As a check, we have shown that the zero-field critical temperature matches the analytical value. Moreover, the behavior of the dAT line for small fields has been shown to obey a very general scaling that has been observed in similar models (see Fig.~\ref{fig:3_2exponent}). 

In the second place, we have shown how to deal with the zero-temperature limit and the corresponding max-sum equations. We have shown that the strong discretization effects that appear when lowering the temperature can be greatly reduced via a smart interpolation procedure. This technique improves a lot the one used previously for the XY model because it is performed in a two-dimensional space. We then applied the max-sum algorithm with the smart interpolation to find the critical field at zero temperature. We have observed a very smooth behavior as a function of the number of grid points $N_p$ and exponentially fast convergence to the $N_p\to\infty$ limit (see Fig.~\ref{fig:maxsum_fitexp} to \ref{fig:maxsum_extrapolation}). The resulting critical field is in a very good agreement with the extrapolation performed using finite-temperature data (Fig.~\ref{fig:fit_lin}).

Finally, we have considered the large connectivity limit of the model. We have shown how to carry on the expansion and explicitly computed $1/Z$ corrections. We have compared the de Almeida-Thouless line found with this ansatz to the true lines obtained by solving the complete Belief Propagation equations and found excellent agreement (see Fig.~\ref{fig:lines_diffZ}).

While our numerical analysis has been carried out for random regular graphs and fields distributed uniformly of the sphere, we highlight that the same procedures can be applied to study different ensembles of graphs (such as Erdős–Rényi graphs) and fields (like Gaussian ones).

\mycolor{Since our methods make heavy use of the Belief Propagation formalism, they are strictly relevant only for sparse random graphs (in which the loops' sizes diverge in the thermodynamic limit) or fully connected topologies (in which the interaction between the spins is weak). Even though they cannot be directly applied to finite-dimensional topologies (like a cubic lattice in 3 dimensions), they could still provide useful insight for finite-dimensional systems. Indeed, one could try to integrate our methods with the recently introduced $M$-layer technique \cite{altieri2017loop}. Despite the fact that the main results of the $M$-layer technique are analytical \cite{angelini2024bethe}, combining the technique with computational approaches like ours could prove a fruitful line of future research \cite{angelini2020loop}.}

\mycolor{Moreover, other studies have observed important similarities between the physical behavior of models defined on sparse random graphs, and the equivalent models defined on finite-dimensional lattices. For example, in a recent work measuring the low-energy excitations in XY and Heisenberg models on sparse random graphs \cite{franz2024soft} many similarities have been found with a similar study performed on a 3D lattice \cite{baity2015soft}.}

Finally, we point out that a possible follow-up to this paper is using the described algorithms and techniques to study the response of single, well-defined graphs to the addition of a perturbation. This analysis, which could be performed both at finite and, thanks to the max-sum algorithm we have presented, zero temperature, might shed additional light on how this system reacts to excitations, thus granting additional information on the physics of the sparse disordered Heisenberg model.

\section*{Acknowledgments}
We are grateful to Chiara Cammarota for helpful discussions. \mycolorsecond{We also thank Cosimo Lupo for providing helpful resources regarding previous results.} The research has received financial support from ICSC – Italian Research Center on High-Performance Computing, Big Data, and Quantum Computing, funded by the European Union – NextGenerationEU.

\bibliographystyle{unsrt}
\bibliography{myreferences}

\pagebreak

\appendix

\section{\mycolor{Generalization to different topologies and field distributions}}\label{app:diffdegreesandfields}

\mycolor{In the main text we focus on random regular graphs and fields distributed uniformly on the sphere. However, the techniques of Sec.~\ref{sec:methods} can also be applied to models with different topologies of the network (for instance, graphs that mimic the behavior of real-world networks) or diverse distributions of the external field.
In order to do so, one has to change the pseudocodes presented in the main text in the following ways:
\begin{itemize}
    \item to modify the field's distribution, one simply needs to extract the field according to the desired distribution, rather than using a uniform distribution over the sphere;
    \item to consider a network characterized by a degree distribution, instead of extracting $Z-1$ integers in the range $ [ 1, \; \mathcal{N}_\text{pop} ] $, one extracts $d_j$ indices, where $d_j$ is a random number extracted according to the excess degree distribution.
\end{itemize}
We note that if one assumes a Dirac delta function as the field distribution (meaning the field is uniform throughout the system), this corresponds to the scenario described in the Introduction for studying the GT line.
}

\section{Saturation problems when dealing with two populations}\label{app:saturation}
Aside from linearizing the BP equations, stability of a solution can be also checked by introducing a slightly perturbed population, $\hat{\mu}$, such that, for each $i$ and $j$, $\hat{\mu}_{i \to j} = \hat{\mu}_{i \to j} + \epsilon v_{ij}$ and $\epsilon v_{ij}$ is uniformly distributed in the interval [-1,1], and evolving it with the same set of equations as the unperturbed one, $\hat{\nu}$. If the distance $\Delta$ between the two populations grows in time, then the fixed point is unstable, otherwise it is stable.

\begin{figure}[!thbp] 
     \centering
     \begin{subfigure}[b]{0.5\textwidth}
         \centering
         \includegraphics[width=\textwidth]{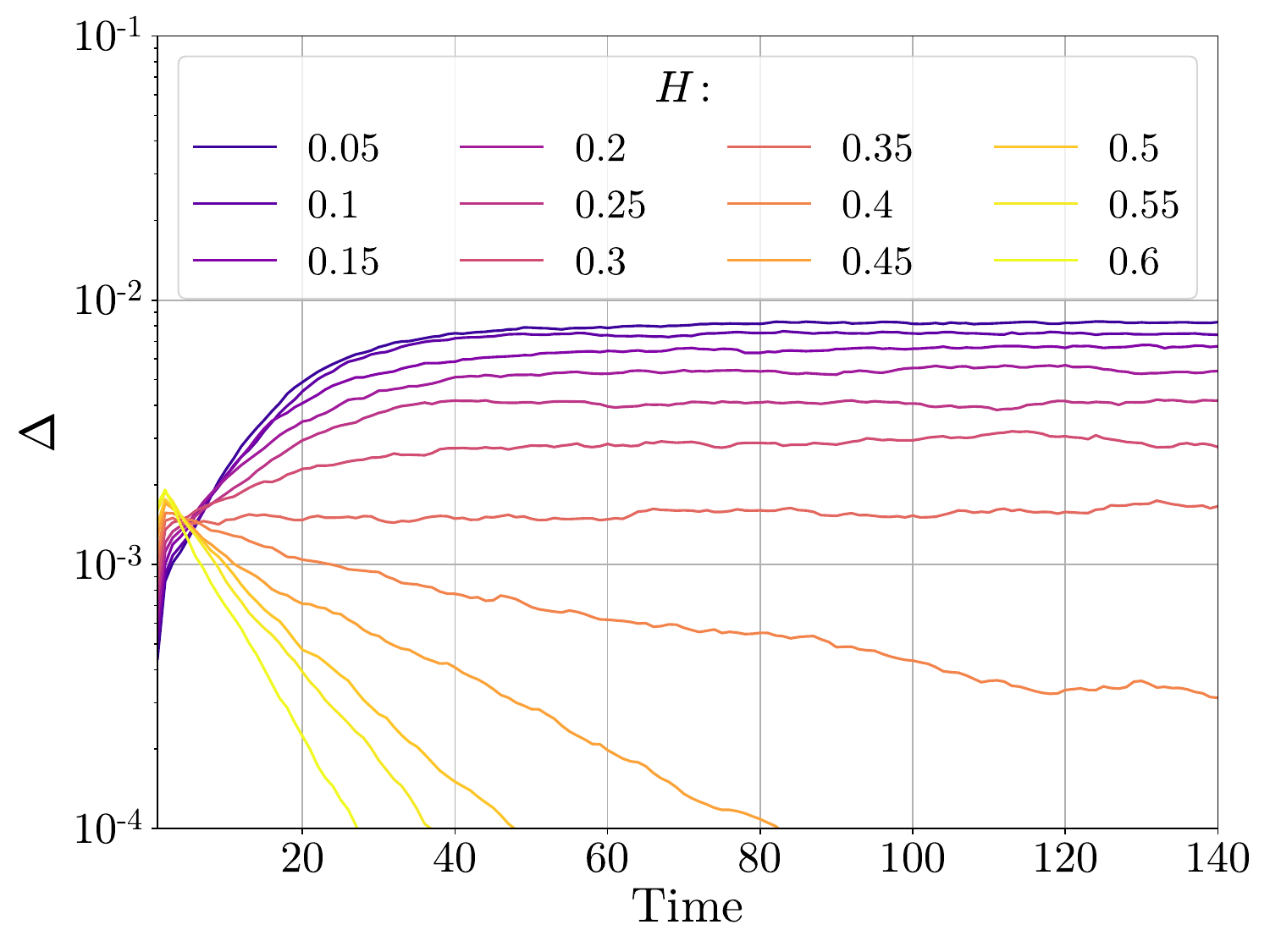}
         \caption{$T = 0.2$.}
         \label{fig:y equals x}
     \end{subfigure}
     \begin{subfigure}[b]{0.5\textwidth}
         \centering
         \includegraphics[width=\textwidth]{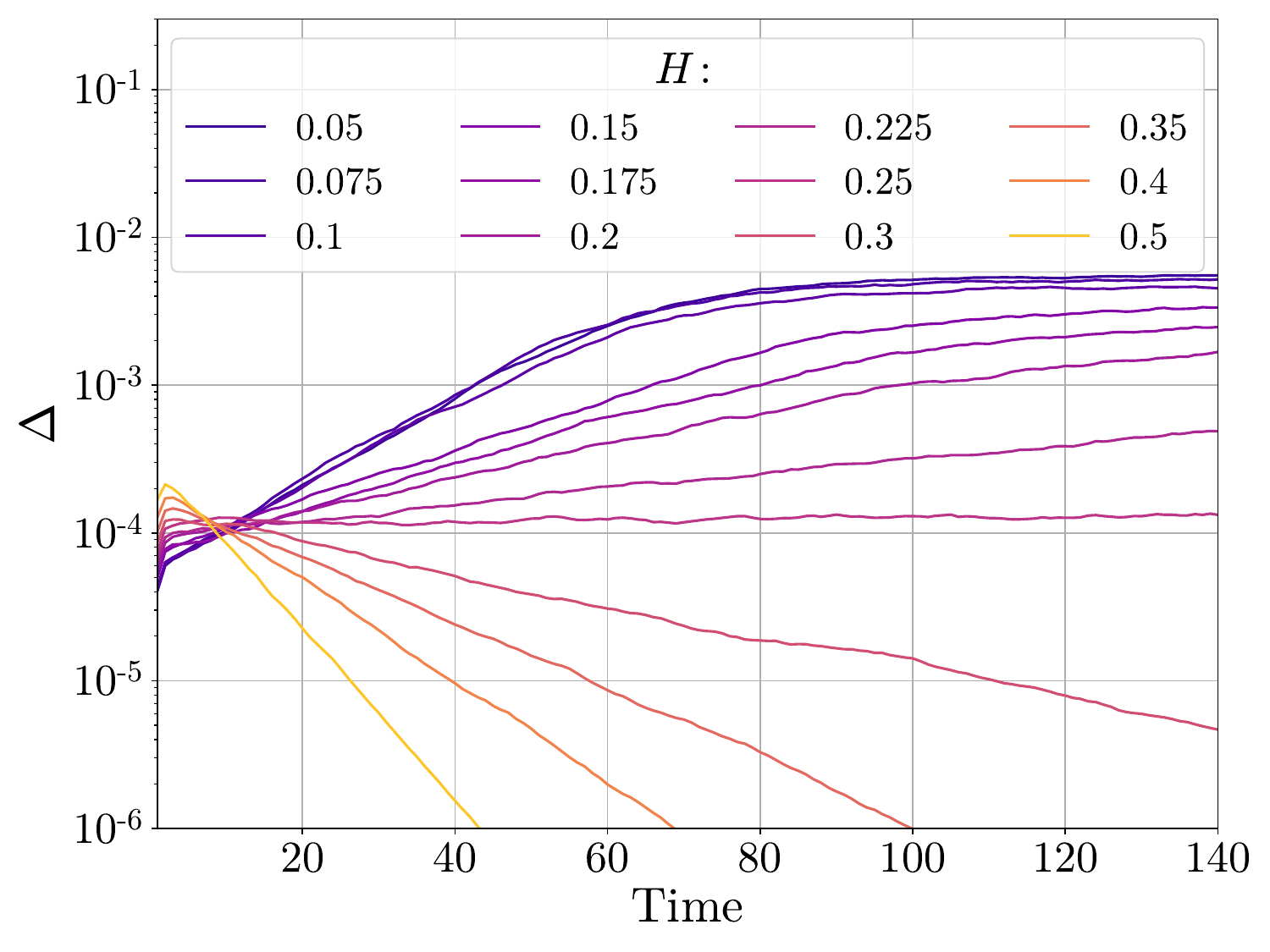}
         \caption{$T = 0.3$.}
         \label{fig:three sin x}
     \end{subfigure}
        \caption[$\Delta$ vs time for the TP algorithm, $\epsilon = 10^{-5}$]{$\Delta$ vs time in semi-log scale for two different values of the temperature. Saturation effects are clearly visible: $\Delta$ stops growing before reaching $10^{-2}$. Parameters: $\epsilon = 10^{-5}$, $t_\text{eq} = 30$, $N_p = 272$ ($V$ potential with $\alpha = 7$).}
        \label{fig:err_series}
\end{figure}

The practical problem of applying this procedure is that, given that marginals are well-normalized on the sphere, the distance between the two populations eventually saturates for long enough times (Fig. \ref{fig:err_series}). If the saturation time is long enough, there is a sufficient number of steps between thermalization and saturation, so that a reliable estimate of the slope, and therefore of the Lyapunov factor, can be carried out nonetheless. However, saturation occurs quicker and quicker for higher fields and lower temperatures, making it impossible to apply the previous procedure to find the critical field in the low-$T$, high-$H$ region. While this problem is temporarily solved by decreasing the initial perturbation, this issue appears again at lower values of the temperature (and higher values of the field), so that one has to continuously decrease the initial perturbation as the temperature is lowered. Since it is not possible to make the initial perturbation small at will due to the finite precision of the computer, a different solution must be used when going to very low temperatures.

\section{\mycolor{A second derivation of the zero-temperature equations for the perturbations}}\label{app:commuting}

\mycolor{In the main text we have derived the equations equations~\eqref{eq:maxsum_deltau} for the evolution of the perturbations $\delta u$ to the cavity fields $u$ starting from the max-sum equations~\eqref{eq:maxsum_u}. That is, we have performed the zero-temperature limit before the linearization. In this appendix, we show that the same equations can be obtained instead by first linearizing the BP equations and then taking the $T \to 0$ limit, thus showing that the two limit commute. Indeed, we have:}
\begin{equation}\label{eq:comm_1}
    e^{\beta \left [ u_{i \to j}(\vec{s}_j) + \delta u_{i \to j}(\vec{s}_j) \right ]} \simeq \mycolorsecond{e^{\beta u_{i \to j}(\vec{s}_j)} \left [1 + \beta \delta u_{i \to j}(\vec{s}_j) \right ]}
\end{equation}
\mycolor{and}
\begin{equation}\label{eq:comm_2}
\mycolor{\begin{gathered}
 e^{\beta \left [ u_{i \to j}(\vec{s}_j) + \delta u_{i \to j}(\vec{s}_j) \right ]} = \frac{\int \text{d} \mu(\vec{s}_i) \; e^{\beta J_{ij}\vec{s}_i\cdot \vec{s}_j} e^{\beta  \vec{H}_i\cdot \vec{s}_i} \prod_{k \in  \partial i \backslash j} e^{\beta \left [ u_{k \to i}(\vec{s}_i) + \delta u_{k \to i}(\vec{s}_i) \right ]}}{\int \text{d} \mu(\vec{s}_j) \int \text{d} \mu(\vec{s}_i) \; e^{\beta J_{ij}\vec{s}_i\cdot \vec{s}_j} e^{\beta  \vec{H}_i\cdot \vec{s}_i} \prod_{k \in  \partial i \backslash j} e^{\beta \left [ u_{k \to i}(\vec{s}_i) + \delta u_{k \to i}(\vec{s}_i) \right ]}}\simeq \\ \simeq \frac{\int \text{d} \mu(\vec{s}_i) \; e^{\beta J_{ij}\vec{s}_i\cdot \vec{s}_j} e^{\beta  \vec{H}_i\cdot \vec{s}_i} \prod_{k \in  \partial i \backslash j} e^{\beta u_{k \to i}(\vec{s}_i)} \left [1 + \beta \delta u_{k \to i}(\vec{s}_i) \right ]}{\int \text{d} \mu(\vec{s}_j) \int \text{d} \mu(\vec{s}_i) \; e^{\beta J_{ij}\vec{s}_i\cdot \vec{s}_j} e^{\beta  \vec{H}_i\cdot \vec{s}_i} \prod_{k \in  \partial i \backslash j} e^{\beta u_{k \to i}(\vec{s}_i)} \left [1 + \beta \delta u_{k \to i}(\vec{s}_i) \right ]}= \\ = \frac{\int \text{d} \mu(\vec{s}_i) \; e^{\beta J_{ij}\vec{s}_i\cdot \vec{s}_j} e^{\beta  \vec{H}_i\cdot \vec{s}_i} e^{\beta \sum_{k \in  \partial i \backslash j} u_{k \to i}(\vec{s}_i)} \left [1 + \beta \sum_{k \in  \partial i \backslash j} \delta u_{k \to i}(\vec{s}_i) \right ]}{\hat{\mathcal{Z}}_{i \to j}(1 + \beta\frac{\delta \mathcal{Z}_{i \to j}}{\mathcal{Z}_{i \to j}})} \simeq \\ \simeq \frac{1}{\hat{\mathcal{Z}}_{i \to j}} \int \text{d} \mu(\vec{s}_i) \; e^{\beta J_{ij}\vec{s}_i\cdot \vec{s}_j} e^{\beta  \vec{H}_i\cdot \vec{s}_i} e^{\beta \sum_{k \in  \partial i \backslash j} u_{k \to i}(\vec{s}_i)} \left (1  + \beta \sum_{k \in  \partial i \backslash j} \delta u_{k \to i}(\vec{s}_i) -  \beta\frac{\delta \mathcal{Z}_{i \to j}}{\mathcal{Z}_{i \to j}}  \right )
 \end{gathered}}
 \end{equation}
\mycolor{and performing the saddle point evaluation for $\beta \to \infty$ and comparing \eqref{eq:comm_1} and \eqref{eq:comm_2} we find that}
\begin{equation}
\mycolor{\delta u_{i \to j}(\vec{s}_j) = \sum_{k \in \partial i \backslash j} \delta u_{k \to i}(\vec{s}_i^{\, *}(\vec{s}_j)) + \text{normalization}}  
\end{equation}
\mycolor{which corresponds to \eqref{eq:maxsum_deltau}.}

\section{Details of the interpolation procedures}
\subsection{Additional complications when dealing with Heisenberg spins with respect to XY spins}\label{app:additional_complications}

Dealing with Heisenberg spins at zero temperature using the max-sum equations introduces the additional following complications with respect to the XY case:

\begin{itemize}
    \item since a truly uniform distribution of points on the sphere cannot be achieved, relative positions of points with respect to each other vary. As a consequence, it is necessary to store, for each of the $N_p$ points of the discretization, the relative position of the neighbours, in order to use them during interpolation;
    \item a 6-parameter interpolation (compared to the 3-parameter one of the XY model) is required. Indeed, considering the space around the discrete argmax locally flat, a second-order polynomial in the $xy$ space has to be considered in order to carry out the interpolation. This means that we need to evaluate the marginals \eqref{eq:maxsum_u} at six different points to find the interpolating polynomial. Since in the grids we are considering (almost) each point has six neighbours, if we take a point and all its neighbours we end up with seven points, so one has to be discarded.\footnote{The alternative would be to perform a fit. However, since the interpolation procedure has to be carried out numerous times at every step of the simulation, it is not computationally feasible to go down this path. Hence, we have chosen to only consider six points and find the (only) paraboloid passing through them.} We have elected to ignore one of the six neighbouring points at random.\footnote{In the case of defects, which only have five neighbours, the choice of the six points is straightforward.} Again, this is unlike the XY case, in which the choice of the three points to perform the interpolation is obvious;
    \item quadratics in two dimensions can have not only minima and maxima but also saddle points. As a consequence, when we try to maximize the function by taking the zero-derivative point there is no guarantee that it will actually be a maximum, but it might very well be a saddle point that lies below the discretized argmax. Moreover, even when the zero-derivative point is actually a maximum, it may lay very far away from the interpolation points. Notice that these issues do not arise in the XY case, in which only maxima are present, and all the maxima lay between the first and the third of the three points considered.
\end{itemize}

\mycolor{In order to overcome these problems during interpolation, we need to adopt a smart interpolation procedure, as explained in the next section.}

\subsection{\mycolor{Maximizing functions on the sphere by using interpolation}}\label{app:max_with_interpolation}

\mycolor{Schematically, we adopt the following procedure to maximize any function $f(\vec{s})$ defined on a sphere:
\begin{itemize}
    \item We find the maximum of $f(\vec{s})$ over the $N_p$ discrete points and we call $\mathsf{P}$ the discrete point where $f(\vec{s})$ reaches its ``discrete'' maximum.
    \item We perform an interpolation with a paraboloid around $\mathsf{P}$. We assume a flat curvature for the sphere, close enough to $\mathsf{P}$, so that we are effectively working in a two-dimensional $xy$ space tangent to the sphere. Therefore the paraboloid is defined by 6 parameters which can be uniquely determined by considering 6 discrete points.
    \item As interpolation points, we consider $\mathsf{P}$ and 5 of its neighbors. If $\mathsf{P}$ has more than 5 neighbours, the 5 points are chosen uniformly at random among the neighbors.
    \item We compute the point of zero gradient for the paraboloid and we call it $\mathsf{Q}$. Since this point could also be a saddle point (and in that case the function in $\mathsf{Q}$ is often lower than in $\mathsf{P}$), we keep $\mathsf{Q}$ as the candidate maximum only if $f(\mathsf{Q}) > f(\mathsf{P})$. Otherwise we use the discrete maximum in $\mathsf{P}$.
    \item Sometimes the interpolating paraboloid has an almost flat curvature and this may bring the point $\mathsf{Q}$ very far from $\mathsf{P}$. This case must be discarded as the true maximum must lie close to $\mathsf{P}$. We only keep a point $\mathsf{Q}$ if its distance from $\mathsf{P}$ is not larger than $\rho r_0$, where $\rho$ is a suitably chosen fixed constant and
    \begin{equation}\label{eq:char_len_0}
        r_0 = \sqrt{\frac{48\eta}{N_p}}
    \end{equation}
    is the average distance between first neighbours. The constant $\eta = \frac{\pi}{\sqrt{12}} \approx 0.91$ is related to the packing density of the 2D hexagonal lattice\cite{chang2010simple}).
\end{itemize}
We still need to choose a suitable value for $\rho$. The details on how to do so are described in the next section.
}
\subsection{Selecting the $\rho$ parameter}\label{app:chosing_rho}

In order to study the effects of $\rho$ on the interpolation procedure, we have first computed numerically the probability distribution $P(x)$ of having the zero-derivative point at a given distance $x$ from the discrete argmax. We expect the distribution to behave almost linearly. Indeed, it is reasonable to assume that the zero-derivative point is distributed uniformly around the discrete argmax, up to half the average distance between neighbours ($x_m$). This leads, for $m \geq 2$, to
\begin{equation} \label{eq:probdist}
    P(x) = C_m x^{m-2}\theta (x_m -x),
\end{equation}
where $C_m$ is a normalization constant that depends on the dimension. For $m = 2$ and $m = 3$ we have
\begin{equation}
    x_2 = \frac{\sqrt{2}\pi}{Q}\;\;\;\;\; C_2 = \frac{1}{x_2},
\end{equation}
\begin{equation}\label{eq:P(x)Heis}
    x_3 = \frac{r_0}{2}\;\;\;\;\; C_3 = \frac{2}{x_3^2},
\end{equation}
respectively. Here, $Q$ is the number of points used to discretize the circle (i.e. the number of \textit{colours} in the clock model).

\begin{figure}[!htbp] 
    \centering
    \makebox[\textwidth][c]{\includegraphics[width=0.7\textwidth]{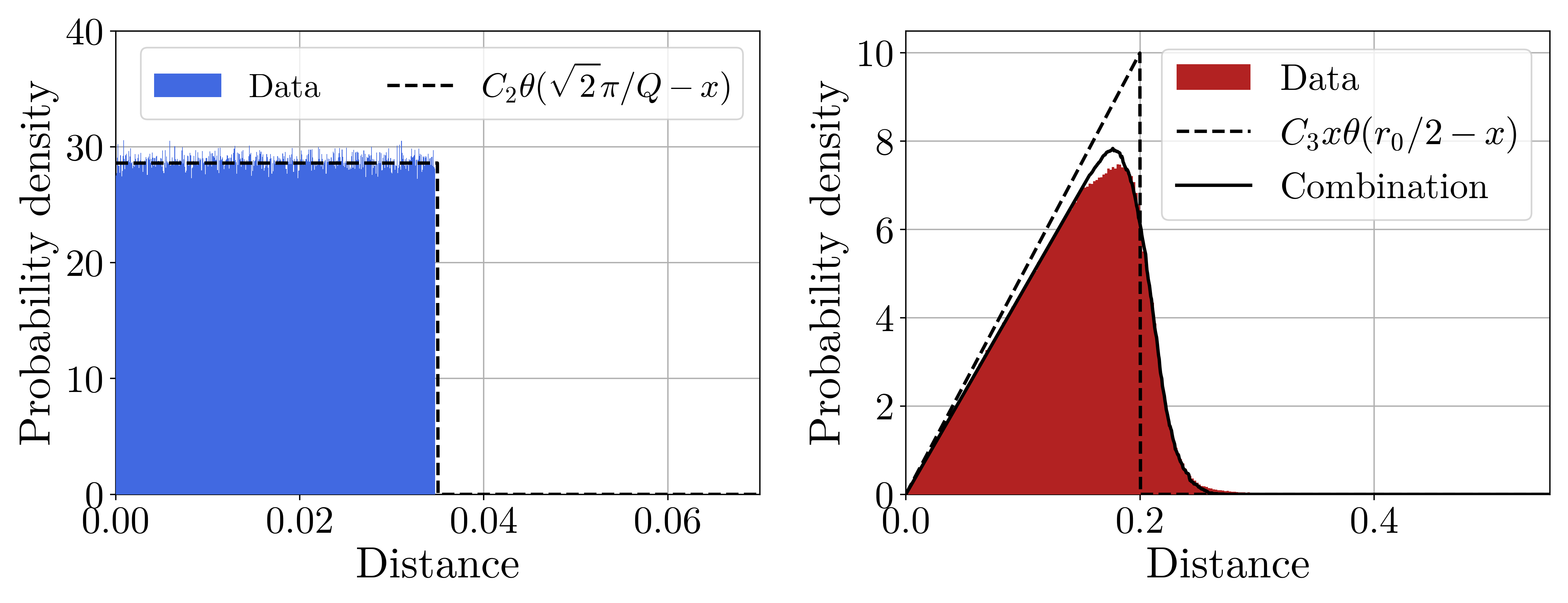}}%
    \caption[Probability distribution of the distance of the zero-derivative point from the discrete argmax]{Probability distribution of the distance $x$ of the zero-derivative point from the discrete argmax. \textit{Left}: XY model with $Q = 128$ points; \textit{right}: the Heisenberg model with $N_p = 272$ ($V$ optimization with $\alpha = 7$) points and $\rho = 1$. For both distributions, the corresponding $P(x)$ has been depicted as a solid line. For the Heisenberg, case an additional prediction has been obtained by combining different distributions in the form \eqref{eq:probdist}, as described in the text. Notice that, in the second case, two different linear regimes are captured in the distribution of the distances. The first is up until the distance reaches the edge of the Voronoi cell, the second is when it reaches its corner.}
    \label{fig:confrontoXYandHeisdistributions}
\end{figure}

We check these hypotheses against data corresponding to the both XY and Heisenberg spins, using the linearized BP algorithm.
Our prediction is clearly verified for the XY model, in which we have a uniform distribution, as reported in the left panel of Fig. \ref{fig:confrontoXYandHeisdistributions}. As is visible in the right panel, the agreement is also satisfactory in the Heisenberg model. Indeed, in this case the starting behaviour is clearly linear, although with a slightly smaller slope than that predicted in \eqref{eq:P(x)Heis}, and a smoothed-out cutoff is present at $x_3 = r_0/2$. The small discrepancies may be due to the fact that the distribution of points on the sphere is not perfectly uniform, so that distances between neighbours are not exactly equal to $x_3$, and moreover defects are present. Additionally, the space surrounding a point should actually be considered a hexagonal Voronoi cell. In order to take (at least partially) into account these effects, we have combined different distributions in the form \eqref{eq:probdist} in which $x_3$ was modified by adding a Gaussian noise (with standard deviation given by the standard deviation of distances between first neighbours) and rescaling everything by $1/\cos \theta$, where $\theta$ is distributed uniformly in the $[-\frac{\pi}{6}, \frac{\pi}{6}]$ interval. The latter procedure has been used to account for, to some extent, the shape of the Voronoi cell. This second prediction appears to resemble closely the empirical distribution of distances.

\begin{figure}[!tbp] 
     \centering
     \begin{subfigure}[b]{0.3\textwidth}
         \centering
         \includegraphics[width=\textwidth]{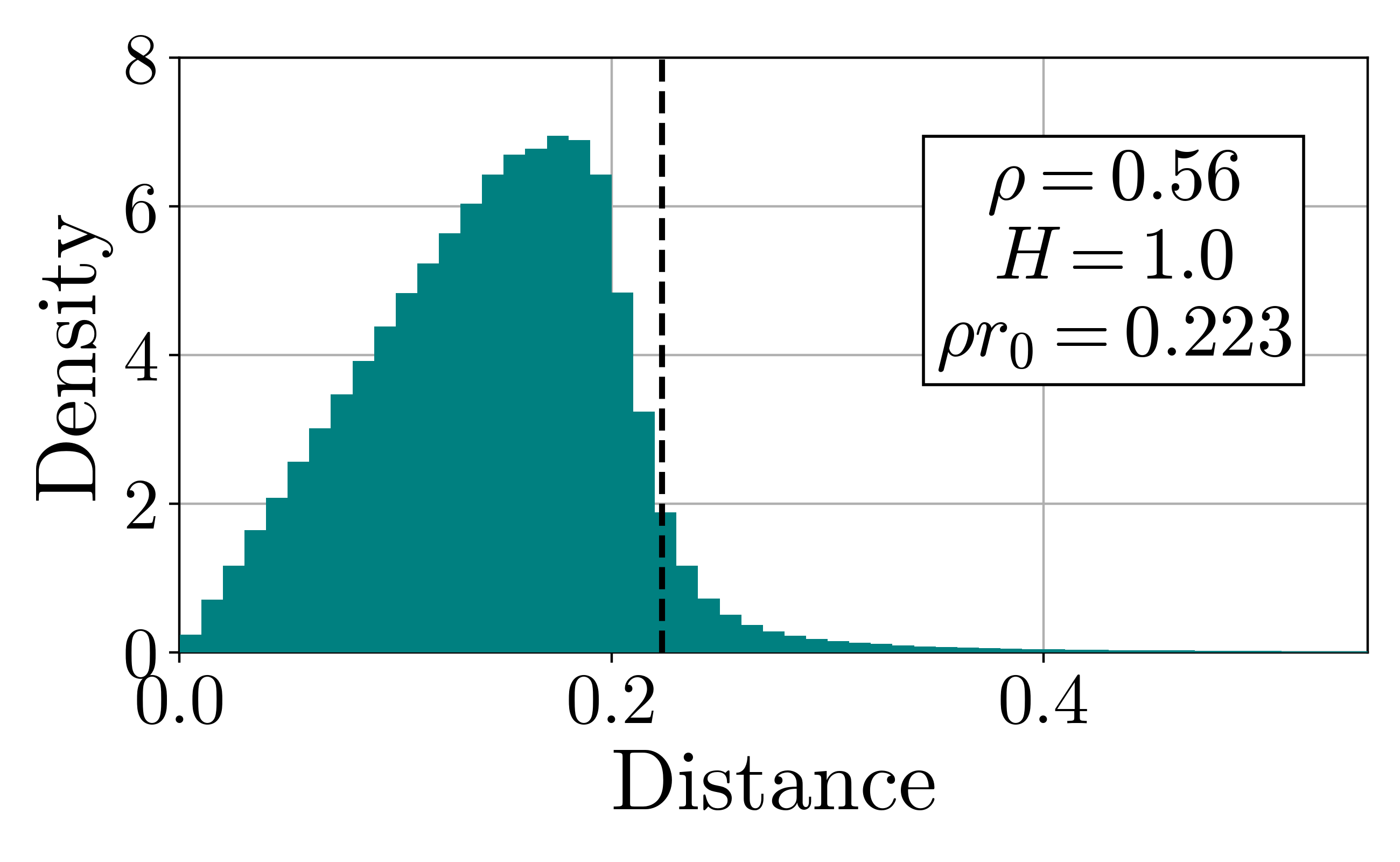}
         \caption{}
         \label{fig:y equals x}
     \end{subfigure}
     \begin{subfigure}[b]{0.3\textwidth}
         \centering
         \includegraphics[width=\textwidth]{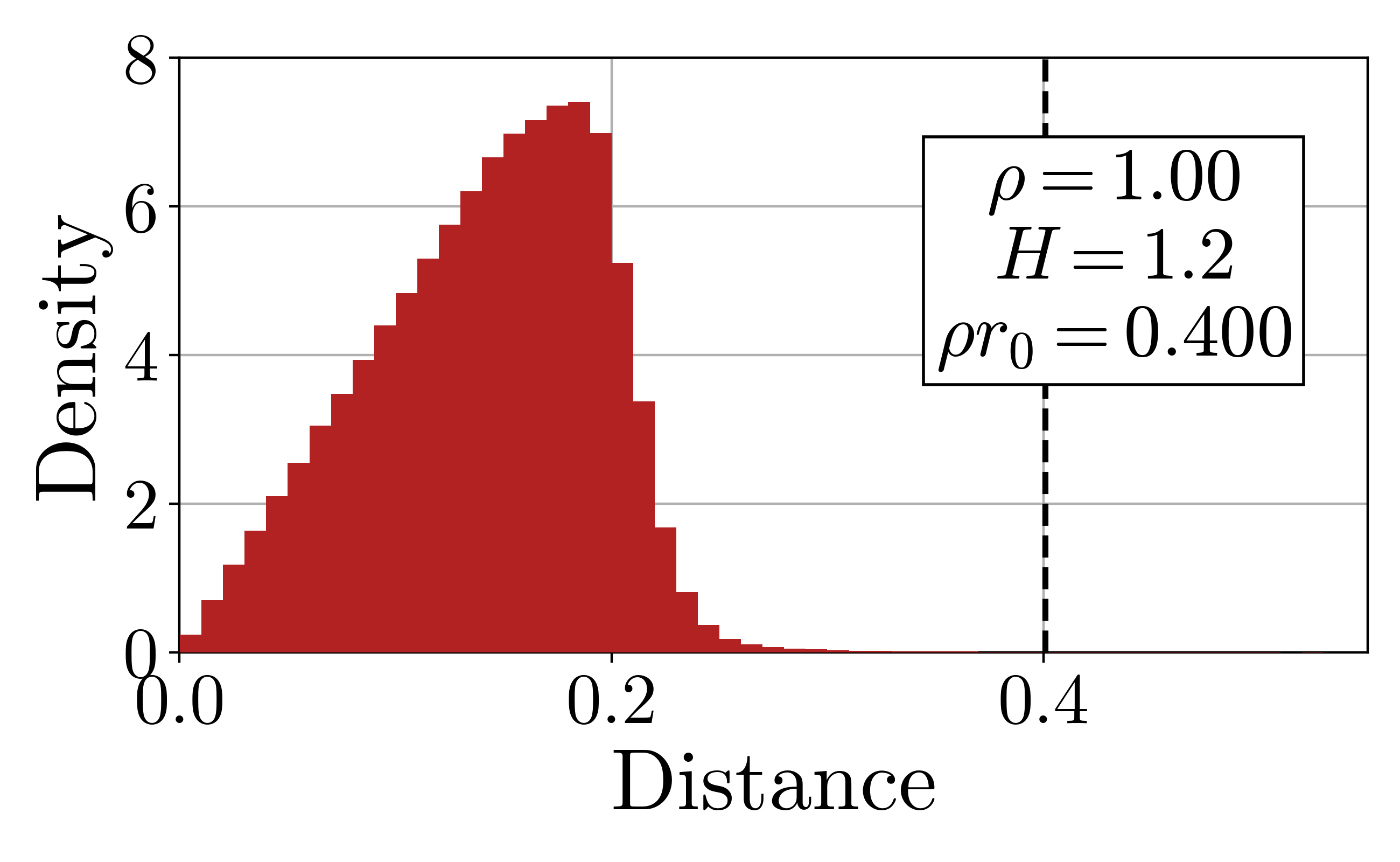}
         \caption{}
         \label{fig:three sin x}
     \end{subfigure}
          \begin{subfigure}[b]{0.3\textwidth}
         \centering
         \includegraphics[width=\textwidth]{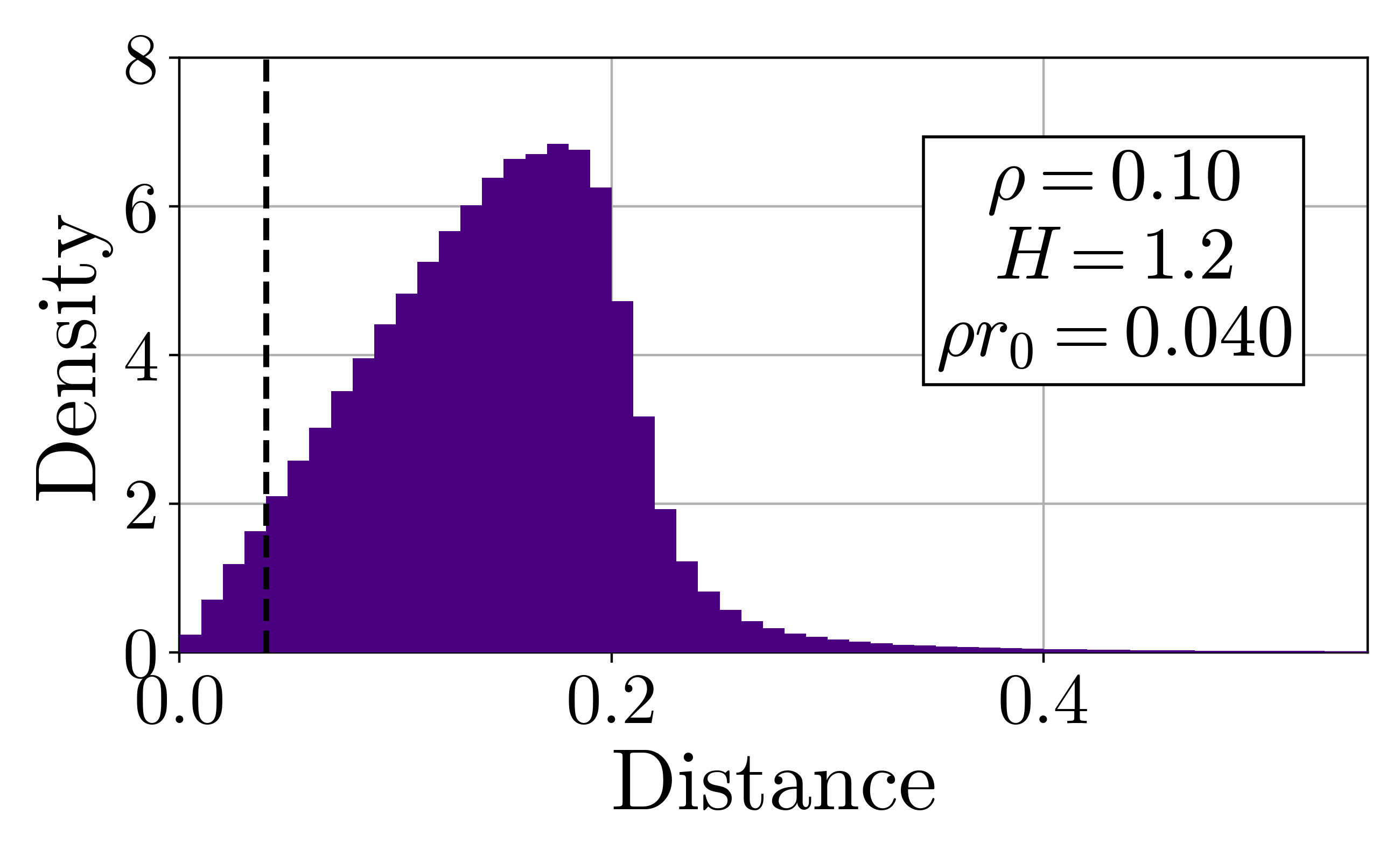}
         \caption{}
         \label{fig:three sin x}
     \end{subfigure}
        \caption[Examples of distribution of the distance from the discrete argmax of the zero-derivative point]{Examples of distribution of the distance from the discrete argmax of the zero-derivative point, for different values of the $\rho$ parameter and of the $H$ field, together with the cutoff $\rho r_0$ for each distribution (dashed line). Data were obtained using $N_p = 272$ ($V$ optimization with $\alpha = 7$) and looking at the 50-th step of the equilibration of the marginals.}
        \label{fig:examples_rhos}
\end{figure}

It can then be investigated how the distribution modifies when $\rho$ changes (Fig. \ref{fig:examples_rhos}). When the cutoff introduced by $\rho$ comes before the natural cutoff of the distribution, a large part of it is discarded. Hence, the algorithm will behave poorly and will not give a correct estimate of the critical field for the continuous model. From this visual analysis alone, values of at least $\rho \gtrsim 0.7$ (corresponding to cutoff distances $\rho r_0 \gtrsim 0.28$) have to be used.

\begin{figure}[!htbp] 
    \centering
    \makebox[\textwidth][c]{\includegraphics[width=0.7\textwidth]{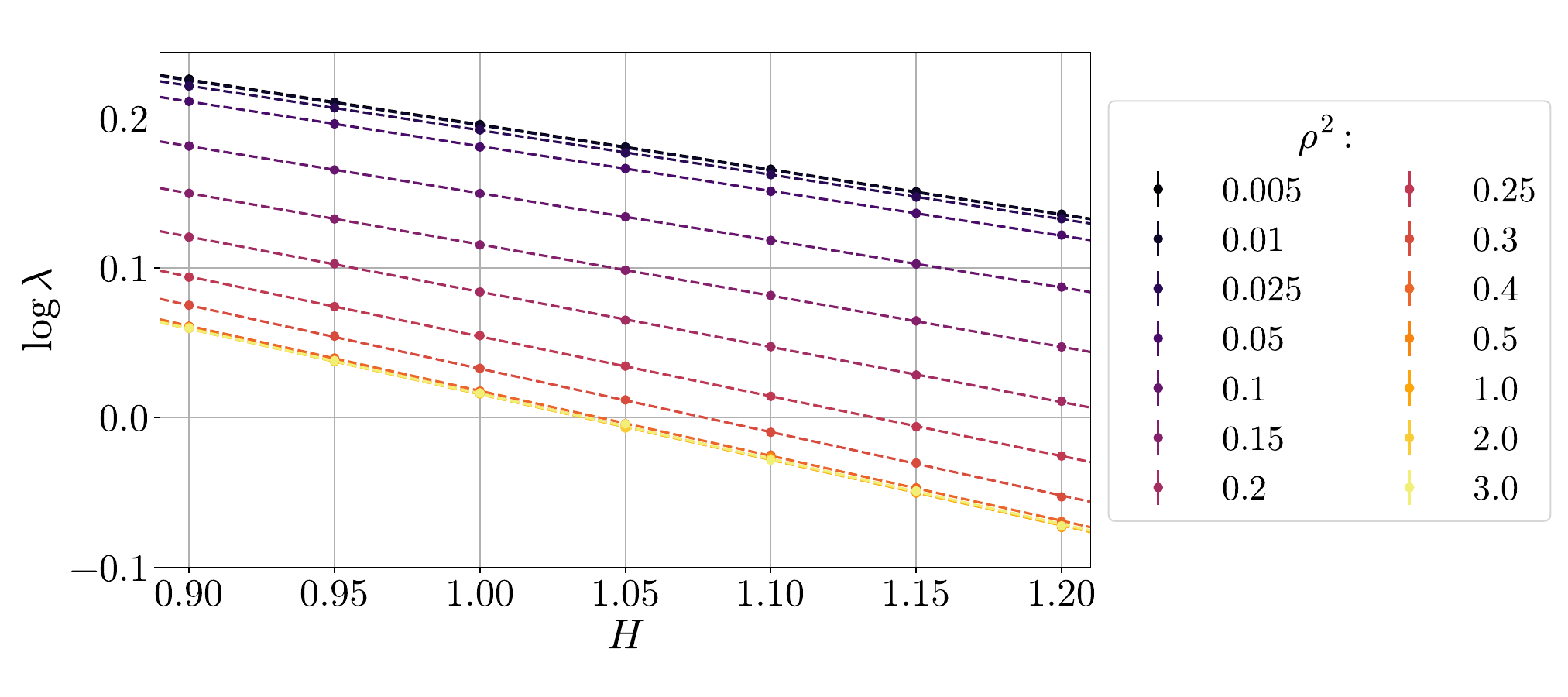}}%
    \caption[Behaviour of $\log \lambda$ vs $H$ for different values of the $\rho$ parameter at $T = 0$]{Behaviour of $\log \lambda$ vs $H$ at $T = 0$ for different values of the $\rho$ parameter. Error bars are smaller than data points. Dashed lines are linear fits. Data for $\rho \leq 0.71$ ($\rho^2 \leq 0.5$) basically collapse onto each other and give the same $H_c(T = 0)$ within errors. All the results have been obtained by averaging together five different runs of $t_\text{tot} = 1000$ steps each ($t_\text{ther} = 100$  steps were excluded to take into account thermalization effects) in order to reduce the noise.}
    \label{fig:logl_diffr}
\end{figure}

\begin{figure}[!htbp] 
    \centering
    \makebox[\textwidth][c]{\includegraphics[width=0.6\textwidth]{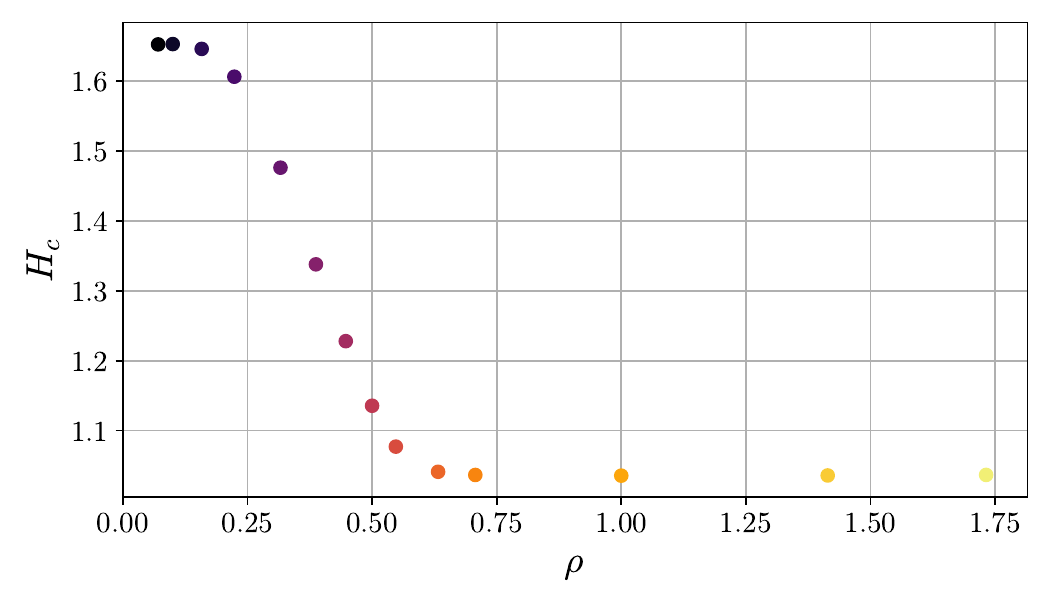}}%
    \caption[Critical field at zero temperature as a function of the $\rho$ parameters]{Critical field at zero temperature, $H_c(T = 0)$, as a function of the $\rho$ parameters. Colours of data points are chosen in order to match those in Fig. \ref{fig:logl_diffr}. Errors are smaller than data points. Notice a smooth transition from the non-interpolating regime at $\rho \gtrsim 0$ to the high-$\rho$ plateau. This difference is the reason why using the max-sum equations without interpolation leads to macroscopically wrong estimates of the critical field.}
    \label{fig:h_vs_r}
\end{figure}

In order to check more quantitatively what happens when $\rho$ is changed, the max-sum algorithm can be used to find, for various $\rho$, the Lyapunov factor at different $H$ and then the value of the critical field at zero temperature. For $N_p = 272$, the values of $\log \lambda$ vs $H$ are shown in Fig. \ref{fig:logl_diffr}. The corresponding $H_c(0)$ as a function of $\rho$ is depicted in Fig. \ref{fig:h_vs_r}. The $H_c$ vs $\rho$ curve has a sigmoidal shape. Clearly, $\rho = 0.71$ ($\rho^2 = 0.5$) already gives a good estimate of the asymptotic value, since we have reached the high-$\rho$ plateau. This result is in agreement with the visual estimate coming from Fig. \ref{fig:examples_rhos}. When $\rho$ reaches unity we are well placed in the plateau, as expected from geometrical considerations (interpolated maxima should not be further away from the discrete maxima than first neighbours), and consequently we can choose $\rho = 1$ as a good value of the parameter for the following simulations. The reader should notice that the $\rho \to 0$ limit (for practical purposes, $\rho \lesssim 0.1$) gives the result of the max-sum algorithm when no interpolation is carried out. This estimate of the zero-temperature critical field is macroscopically wrong.

\section{Computation of the dAT line \mycolor{in the fully connected case} using the vectorial ansatz}\label{app:vectorial_ansatz}

\mycolor{In the following, we show how to locate the dAT line for systems of Heisenberg spins using the vectorial ansatz \cite{javanmard2016phase}. This ansatz, as evident from the computations of this Appendix, is exact in the case of fully connected models, and thus allows us to find the fully-connected dAT line for our model exactly.}

\subsection{General computation}

Starting from \eqref{eq:largeZ_nu} and \eqref{eq:largeZ_nuhat}, the self-consistency equations for the cavity fields and the cavity magnetizations become
\bea
\bh_{i\to j} &=& \bH_i + \sum_{k\notin\{i,j\}} J_{ik} \bmm_{k\to i}
\eea
\bea \label{eq:FCmagnetizations}
\bmm_{i\to j} &=& \frac{\int \dd \mu (\bs_i) \exp(\beta \bh_{i\to j} \cdot \bs_i) \bs_i}{\int \dd \mu (\bs_i) \exp(\beta \bh_{i\to j} \cdot \bs_i)}
= \sqrt{3} f_3(\sqrt{3} \beta |\bh_{i\to j}|) \frac{\bh_{i\to j}}{|\bh_{i\to j}|}
\eea
thanks to the symmetry of the numerator's integrand around $\bh_{i\to j}$. Here,
\bea
\label{eq:langevin_function}
f_3(x) = \frac{1}{\tanh(x)}-\frac{1}{x}
\eea
and the $\sqrt{3}$ factor comes from the normalization of the spins.
For completeness we also write down the Ising and XY cases: $f_1(x)=\tanh(x)$ and $f_2(x)=I_1(x)/I_0(x)$. Moreover, $f_m(x)\simeq 1-(m-1)/(2x)$ for $x\to\infty$ for any dimension $m$ of the variables, a fact that will be used later when dealing with the zero-temperature limit.



Perturbations around the BP fixed point can now be studied by linearizing the previous equations. Indeed, we can consider the susceptibility (Greek indices run over coordinates)
\bea
\chi_{\alpha \gamma} \equiv  \frac{\dd m_\alpha}{\dd h_\gamma} = \sqrt{3}\frac{ f_3(\sqrt{3} \beta h)}{h} \delta_{\alpha,\gamma} + \left( 3 \beta f_3^\prime(\sqrt{3} \beta h) - \sqrt{3} \frac{f_3(\sqrt{3} \beta h)}{h}\right) \frac{h_\alpha h_\gamma}{h^2},
\eea
where the $i \to j$ subscripts have been omitted and $h = |\vec{h}|$.
Calling $\bn_{i\to j} = \bh_{i\to j} / |\bh_{i\to j}|$, the susceptibility matrix can then be written as
\bea
\bm{\chi}_{i\to j} = \sqrt{3} \frac{f_3(\sqrt{3} \beta h_{i\to j})}{h_{i\to j}} \mathds{1} + \left(3 \beta f_3^\prime(\sqrt{3}\beta h_{i\to j}) -  \sqrt{3}\frac{f_3(\sqrt{3}\beta h_{i\to j})}{h_{i\to j}}\right) |\bn_{i\to j}\>\<\bn_{i\to j}|
\eea
$\mathds{1}$ being the identity matrix. Linear perturbations then evolve according to the random matrix 
\bea
\bm{B}_{i\to j} = \sum_{k\notin\{i,j\}} J_{ik} \bm{\chi}_{k\to i},
\eea
and this implies that the covariance of their distribution $\bm{C}$ evolves according to $\bm{C}^{t+1} = \bm{B}\bm{C}^t\bm{B}$.
Assuming that the initial perturbations are randomly distributed according to a normal distribution of zero mean and covariance $\1$ \mycolor{and considering random independent couplings extracted according to a distribution of zero mean and variance one}, their evolution can be readily obtained from the study of
\bea
\bm{B}^2_{i \to j} \simeq \E[\bm{\chi} ^2_{i \to j}] = \E\left[2 \left(\frac{f_3(\sqrt{3} \beta h)}{h}\right)^2 + 3 \left(\beta f_3^\prime(\sqrt{3} \beta h)\right)^2 \right] \1,
\eea
where we have used $\E\left[|\bn\>\<\bn|\right]=\mathds{1}/3$. The average is over the probability distribution of the norm of $\vec{h}_{i \to j}$. Finally, the dAT line is defined by the condition
\bea \label{eq:datLineEQ}
\E_h\left[2 \left(\frac{f_3(\sqrt{3} \beta h)}{h}\right)^2 + 3\left(\beta f_3^\prime(\sqrt{3} \beta h)\right)^2 \right] = 1.
\eea
Equation \eqref{eq:datLineEQ} is equivalent to the condition $\lambda = 1$ encountered in the main text and it is easy to check that \mycolor{it} provides the same dAT line computed in \cite{sharma_and_young}\mycolor{, equations (71) and (C26), once one considers external Gaussian fields, as shown at the beginning of the next section}.

\subsection{Solving the stability equations at zero temperature}

\subsubsection{Gaussian fields}

Let us consider \mycolor{first} the case in which the components of the field are distributed as Gaussian variables of zero mean and variance $H^2$. 
Then, by the central limit theorem, cavity fields $\bh_{i\to j}$ are Gaussian random variables of zero mean and covariance matrix $\sigma_h^2\mathds{1}$, with $\sigma_h^2$ satisfying
\bea
\sigma_h^2 &=& H^2 + \frac13 \mathbb{E}[\bmm^2] = H^2 +  \E[f_3(\sqrt{3} \beta h)^2] =\nonumber\\
&=& H^2 + \frac{\int_0^\infty dh\,h^2\,e^{-h^2/(2\sigma_h^2)} f_3( \sqrt{3} \beta h)^2}{\int_0^\infty dh\,h^2\,e^{-h^2/(2\sigma_h^2)}}.
\eea
In the zero temperature limit the equation reduces to $\sigma_h^2=H^2+1$. The average appearing in \eqref{eq:datLineEQ} then has the form
\bea
\E[\bullet] = \sqrt{\frac{2}{\pi}} \int_0^\infty \bullet\, \frac{dh\,h^2}{\sigma_h^3} e^{-\frac{h^2}{2\sigma_h^2}}
\eea
and, in the zero temperature limit, the dAT condition becomes
\bea \label{eq:gauss_analytical}
1 = 2 \sqrt{\frac{2}{\pi}} \int_0^\infty \frac{dh\,e^{-\frac{h^2}{2\sigma_h^2}}}{\sigma_h^3} = \frac{2}{\sigma_h^2} \implies H_c(T = 0) =  1.
\eea
In general, for $m$-dimensional variables the critical field is at $H_c(T = 0) = 1/\sqrt{m-2}$, in complete agreement with the result in \cite{sharma_and_young}. Notice that the zero-temperature critical field diverges for $m \leq 2$.

\subsubsection{Uniformly distributed fields}

Now, let us move to the case \mycolor{considered in the main text,} in which fields are distributed uniformly on the sphere of radius $H$. Due to isotropy, equations are invariant under the rotation of the local field $\vec{H}$, so we can fix its direction suitably. For instance, we can consider it aligned in the direction $\hat{n} = (1,1,1)$, so that (again, we drop the $i \to j$ subscripts to simplify the notation)
\begin{align}\label{eq:h_porcedure1}
\begin{split}
    h_1 = \frac{H}{\sqrt{3}} + \sigma z_1 \\
    h_2 = \frac{H}{\sqrt{3}} + \sigma z_2 \\
    h_3 = \frac{H}{\sqrt{3}} + \sigma z_3 \\
\end{split}
\end{align}
and $z_\alpha$, $\alpha = 1,2,3$ are normal variables of zero mean an variance one. Due to symmetry, the standard deviation $\sigma$ is the same for all components. Since the variance of the field is zero in this case, $\sigma$ has to satisfy 
\begin{equation}
    \sigma^2 = \frac{1}{3} \mathbb E [\vec{m}^2].
\end{equation}
From \eqref{eq:FCmagnetizations} with $\beta \to \infty$,
\begin{equation}
    (m_1)^2 + (m_2)^2 +(m_3)^2 = 3
\end{equation}
and the fact that the three components are equal, it follows that
\begin{equation}
    \sigma = 1.
\end{equation}
Then, the expected value in \eqref{eq:datLineEQ} can be easily computed by generating the components of $\vec{h}$ as Gaussian variables of mean $H/\sqrt{3}$ and standard deviation 1. 

Alternatively, we can align the external field to the $\hat{n} = \hat{x} = (1,0,0)$ direction so that
\begin{equation}
\begin{gathered}
    h_1 = H + \sigma z_1 \\
    h_2 =  \sigma z_2 \\
    h_3 = \sigma z_3 \\
\end{gathered}
\end{equation}
where $\sigma = 1$ as before.
Then again, we can compute the expected value in \eqref{eq:datLineEQ} numerically for different values of the field, obtaining the curve of the expected value as a function of $H$.

\begin{figure}[htb]
    \centering
    \makebox[\textwidth][c]{\includegraphics[width=0.6\textwidth]{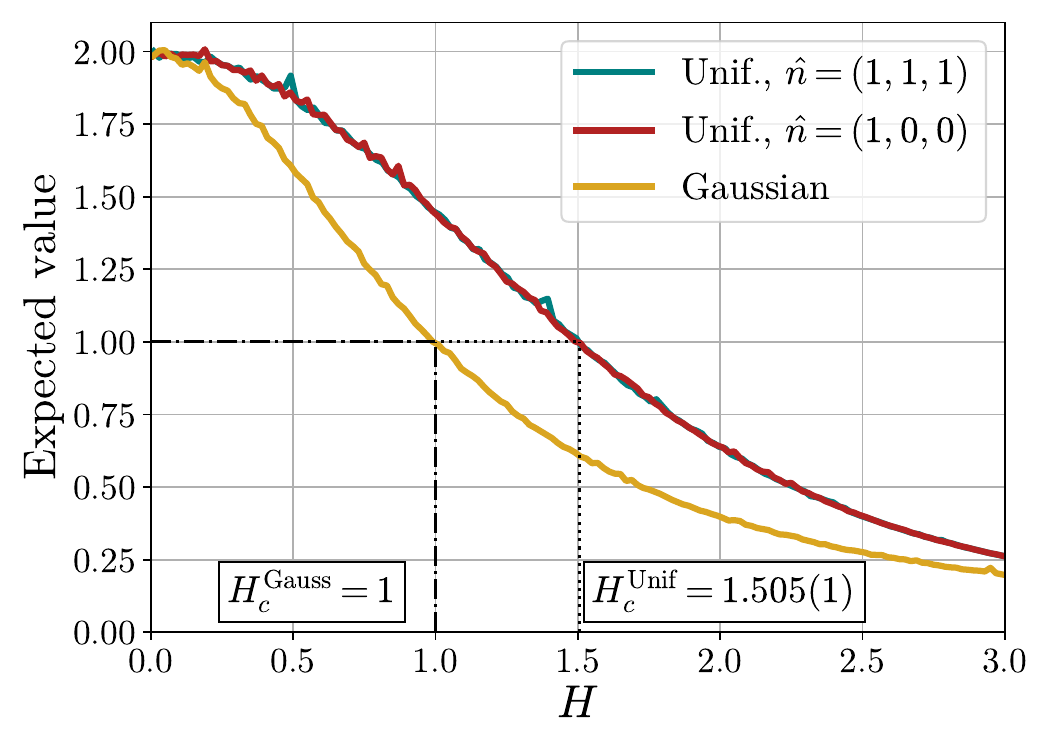}}%
    \caption[Comparison of the FC stability lines at $T = 0$ for  Gaussian and uniformly distributed fields]{Comparison of the FC lines of the expected value in \eqref{eq:datLineEQ} at $T = 0$, for  Gaussian and uniformly distributed fields. The critical field is identified by the reaching of unity. The two curves obtained aligning the field in the $(1,0,0)$ and $(1,1,1)$ directions (see main text) coincide.}
    \label{fig:lines_FC_0T}
\end{figure}

The comparison between the lines at zero temperature for Gaussian and uniformly distributed fields is carried out in Fig. \ref{fig:lines_FC_0T}, in which the agreement between the two different methods described above for  the uniform-distribution case can also be appreciated.

In the end, the zero-temperature critical fields are
\begin{equation}
    H_c^\mathrm{Gauss}(T = 0) = 1,
\end{equation}
\begin{equation}
    H_c^\mathrm{Unif}(T = 0) = 1.505 \pm 0.001,
\end{equation}
respectively. The first result derives from the analytical computation \eqref{eq:gauss_analytical}; the second one comes from a linear fit  of the curves presented in Fig. \ref{fig:lines_FC_0T} in the region of interest.

\subsection{Solving the stability equation at finite temperature}

When going to $T > 0$ \eqref{eq:datLineEQ} can still be solved using fundamentally the same procedure used when dealing with the zero-temperature limit. Indeed, we can still use, for example, the first method described in the previous section in order to write the expected value in \eqref{eq:datLineEQ}. The main difference with the $T = 0$ case is that now $f_3$ is not identically equal to one, so $\sigma \neq 1$ and its value is not known from the beginning. The exact value of $\sigma$ can, however, be found as $\sigma^2 = \mathbb{E}[\bmm^2] / 3$, where now
\begin{equation}\label{eq:m_for_sigma}
    \vec{m}^2 = 3 f_3(\sqrt{3} \beta h)^2
\end{equation}
from \eqref{eq:FCmagnetizations}. Hence, for a given temperature $T$, we can set up a recursive procedure, using $\sigma$ to evaluate, via \eqref{eq:h_porcedure1}, $\vec{h}$ (and therefore $h = |\vec{h}|$) and then using the latter to compute first $\vec{m}^2$ using \eqref{eq:m_for_sigma} and then a new value of $\sigma$. The value to which the process converges is the estimate of $\sigma$, that can then be used to compute the expected value in \eqref{eq:datLineEQ}. As in the $T = 0$ case, the procedure can be applied at different field intensities $H$ and the value $H_c(T)$ at which the expected value reaches one signals the onset of the spin glass phase.

The process can be repeated at different temperatures in order to compute the whole dAT line. Results can then be checked against those obtained at finite $Z$ in order to verify that the former are obtained in the  $Z \to \infty$ case, as done in Fig. \ref{fig:lines_diffZ}. Despite some differences, in particular in the low-$T$ region, the convergence to the FC case can be clearly appreciated for increasing values of the connectivity, as evidenced by the fact that the data points move closer to the FC solution for larger values of $Z$.

\section{Quartic expansion of the BP equations}\label{app:expansion}

\mycolor{When moving from the fully connected limit to the finite-connectivity case, the vectorial ansatz used in App. \ref{app:vectorial_ansatz} is no longer exact. One can, however, turn into an approximation that can be improved by considering higher order terms. Indeed, } starting from \eqref{eq:BPnuhat} and expanding the exponential under the assumption that $\beta J_{ij}$ is small we find, up to the fourth order:
\begin{equation}
    \hat{\nu}_{j \to i}(\vec{s}_i) \propto \int \text{d} \mu(\vec{s}_j) \; \left [1 + \beta J_{ij}\vec{s}_i\cdot \vec{s}_j + \frac{1}{2}(\beta J_{ij}\vec{s}_i\cdot \vec{s}_j)^2 + \frac{1}{3!} (\beta J_{ij}\vec{s}_i\cdot \vec{s}_j)^3 + \frac{1}{4!} (\beta J_{ij}\vec{s}_i\cdot \vec{s}_j)^4 \right ] \nu_{j \to i}(\vec{s}_j),
\end{equation}
which, once we have defined the correlation functions of $\nu_{j \to i}$,
\begin{equation}\label{eq:par1_start}
    m_{i \to j}^{\alpha} = \int \text{d} \mu(\vec{s}_j) \nu_{j \to i}(\vec{s}_j) s_j^\alpha
\end{equation}
\begin{equation}
    C_{i \to j}^{\alpha \beta} = \int \text{d} \mu(\vec{s}_j) \nu_{j \to i}(\vec{s}_j) s_j^\alpha s_j^\beta
\end{equation}
\begin{equation}
    T_{3, i \to j}^{\alpha \beta \gamma} = \int \text{d} \mu(\vec{s}_j) \nu_{j \to i}(\vec{s}_j) s_j^\alpha s_j^\beta s_j^\gamma
\end{equation}
\begin{equation}\label{eq:par1_end}
    T_{4, i \to j}^{\alpha \beta \gamma \delta} = \int \text{d} \mu(\vec{s}_j) \nu_{j \to i}(\vec{s}_j) s_j^\alpha s_j^\beta s_j^\gamma s_j^\delta
\end{equation}
becomes:
\begin{equation}
    \hat{\nu}_{j \to i}(\vec{s}_i) \propto 1 + \beta J_{ij} \sum_{\alpha} m_{i \to j}^{\alpha} s_i^\alpha +
    \frac{(\beta J_{ij})^2}{2}\sum_{\alpha \beta} C_{i \to j}^{\alpha \beta} s_i^\alpha s_i^\beta +
    \frac{(\beta J_{ij})^3}{3!} \sum_{\alpha \beta \gamma} T_{3, i \to j}^{\alpha \beta \gamma} s_i^\alpha s_i^\beta s_i^\gamma +
    \frac{(\beta J_{ij})^4}{4!}\sum_{\alpha \beta \gamma \delta} T_{4, i \to j}^{\alpha \beta \gamma \delta} s_i^\alpha s_i^\beta s_i^\gamma s_i^\delta
\end{equation}
Writing $\hat{\nu}_{j \to i}$ as the exponential of the logarithm and, again, expanding for small $\beta J_{ij}$ yields the quartic approximation of the marginals
\begin{equation}
    \hat{\nu}_{j \to i}(\vec{s}_i) \propto \exp \left \{ 
    \sum_{\alpha} \Tilde{m}_{i \to j}^{\alpha} s_i^\alpha +
    \sum_{\alpha \beta} \Tilde{C}_{i \to j}^{\alpha \beta} s_i^\alpha s_i^\beta +
    \sum_{\alpha \beta \gamma} \Tilde{T}_{3, i \to j}^{\alpha \beta \gamma} s_i^\alpha s_i^\beta s_i^\gamma +
    \sum_{\alpha \beta \gamma \delta} \Tilde{T}_{4, i \to j}^{\alpha \beta \gamma \delta} s_i^\alpha s_i^\beta s_i^\gamma s_i^\delta
    \right \}
\end{equation}
where
\begin{equation}\label{eq:par2_start}
    \Tilde{m}_{i \to j}^{\alpha} = \beta J_{ij} m_{i \to j}^{\alpha}
\end{equation}
\begin{equation}
    \Tilde{C}_{i \to j}^{\alpha \beta} = \frac{(\beta J_{ij})^2}{2} \left ( C_{i \to j}^{\alpha \beta} -  m_{i \to j}^{\alpha} m_{i \to j}^{\beta}\right )
\end{equation}
\begin{equation}
    \Tilde{T}_{3, i \to j}^{\alpha \beta \gamma} = (\beta J_{ij})^3 \left ( \frac{1}{6}T_{3, i \to j}^{\alpha \beta \gamma} - \frac{1}{2} C_{i \to j}^{\alpha \beta} m_{i \to j}^{\gamma} + \frac{1}{3} m_{i \to j}^{\alpha} m_{i \to j}^{\beta} m_{i \to j}^{\gamma}  \right )
\end{equation}
\begin{equation}\label{eq:par2_end}
    \Tilde{T}_{4, i \to j}^{\alpha \beta \gamma \delta} = (\beta J_{ij})^4 \left ( \frac{1}{24} T_{4, i \to j}^{\alpha \beta \gamma \delta} -\frac{1}{8} C_{i \to j}^{\alpha \beta} C_{i \to j}^{\gamma \delta} + \frac{1}{2} C_{i \to j}^{\alpha \beta} m_{i \to j}^{\gamma} m_{i \to j}^{\delta}-\frac{1}{4} T_{3, i \to j}^{\alpha \beta \gamma} m_{i \to j}^{\delta} -\frac{1}{6} m_{i \to j}^{\alpha} m_{i \to j}^{\beta} m_{i \to j}^{\gamma} m_{i \to j}^{\delta}
    \right )
\end{equation}
This form is preserved for $\nu_{j \to i}$, from \eqref{eq:BPnu}.
Therefore, starting from $\nu_{j \to i}$, one can find the values of the parameters \eqref{eq:par1_start}-\eqref{eq:par1_end}, than those of \eqref{eq:par2_start}-\eqref{eq:par2_end} and from these compute again the $\nu_{j \to i}$ from \eqref{eq:BPnu}. The procedure can then be repeated until convergence is reached. Notice that, if one is interested only in results up to $\mathcal{O}(1/Z)$, as we are in this paper, one only has to consider positive elements of $\Tilde{T}_{4, i \to j}^{\alpha \beta \gamma \delta}$, that is, elements for which indices are either all equal or pairwise equal.

\section{Zero temperature extrapolation from finite-temperature data}\label{app:extrapolation}

\begin{figure}[!htbp] 
    \centering
    \makebox[\textwidth][c]{\includegraphics[width=0.6\textwidth]{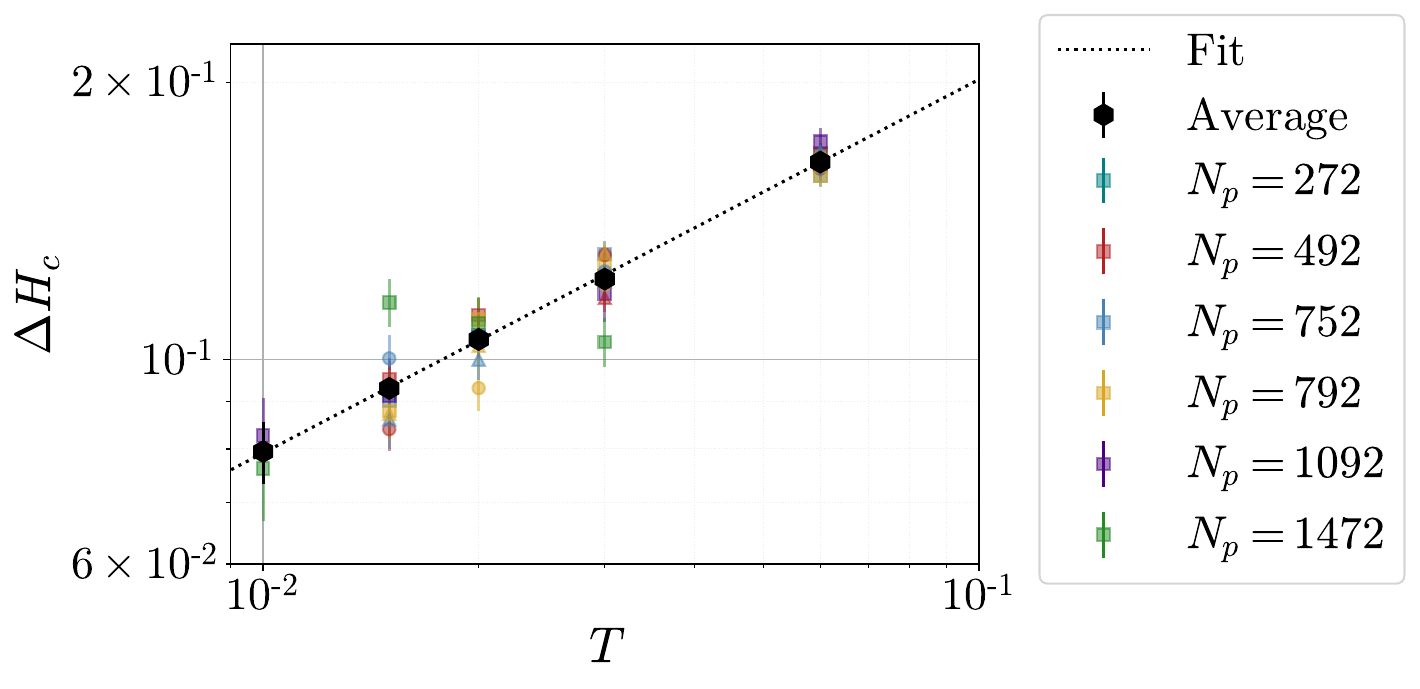}}%
    \caption[$\Delta H_c(T)$ vs $T$ for different numbers of points $N_p$]{$\Delta H_c(T)$ vs $T$ for different numbers of points $N_p$ in semi-log scale. Data were obtained using the linearized BP algorithm. Multiple points with the same colour (that is, with the same $N_p$) correspond to different values of $\alpha = 1,4,7$ in the choice of the optimization potential for the grid $V(r)_\alpha \propto r^{-\alpha}$ (see \cite{delebono2024uniform} for details). \text{Squares}: LP grids optimized with $V$ potential with $\alpha = 1$; \text{circles}: LP grids optimized with $V$ potential with $\alpha = 4$; \text{triangles}: LP grids optimized with $V$ potential with $\alpha = 7$. Black hexagons are averages, and the dotted line is the result of the fit.}
    \label{fig:fitalpha}
\end{figure}

The procedure used to estimate the critical field at zero temperature from the finite-temperature data is the following. First, we exclude $H_c(0)$ from the picture in order to reduce the number of fitting parameters. This is done by assuming a power-law behaviour at low temperatures in the form
\begin{equation}\label{eq:powerlawbehaviour}
    H_c(T) \simeq H_c(0) - AT^a,
\end{equation}
we can consider
\begin{equation}
    \Delta H_c(T) \equiv H_c(T) - H_c(2T),
\end{equation}
which then, according to \eqref{eq:powerlawbehaviour}, goes as
\begin{equation}
    \Delta H_c(T) \simeq AT^a(2^a-1).
\end{equation}
Taking the logarithm yields:
\begin{equation}
    \log \Delta H_c(T) \simeq a \log T + \text{constant},
\end{equation}
so via a linear fit in log-log scale we can obtain the value of $a$. The procedure is depicted in Fig. \ref{fig:fitalpha} and the corresponding value $a$ found in this way is
\begin{equation}
    a = 0.405 \pm 0.014.
\end{equation}
Notice that, in order to have a reliable estimate of the exponent $a$, the points affected by finite size effects must be ignored.

The fitting procedure is in good accord with the measurements. Now we can go back to \eqref{eq:powerlawbehaviour} and fit the data using the known value of $a$ (Fig. \ref{fig:fit_lin}). This procedure yields
\begin{equation}
    H_c(0) = 1.20 \pm 0.02
\end{equation}
as our estimate of the zero temperature critical field. The second fit, i.e. the one that determines $H_c(0)$ starting from the values of $A$ and $a$, depends on the results of the first one. Consequently, it is affected by their uncertainties. In order to take into account this effect,  the error is given by looking at the difference between the value of $H_c$ obtained using the best estimates of $A$ and $a$ and the extreme values obtained by varying $A$ and $a$ within the range defined by their errors.

\section{The equation for the zero field critical temperature}\label{app:_risultati_articolo}

In \cite{coolen2005finitely} it is shown that the critical temperature of vector spin glasses in absence of an external magnetic field is obtained by requiring that the maximal eigenvalue of the following kernel\mycolor{, given in equation (62) of \cite{coolen2005finitely},}
\begin{equation}
\label{eq:kernel_coolen}
    \mathcal{K}(\bs_1, \bs_2, \vec{t}_1, \vec{t}_2)\,=\,\frac{B\exp\left[\beta J\left(\bs_1\cdot\vec{t}_1+\bs_2\cdot \vec{t}_2\right) \right]}{\int \mycolor{\dd \tilde{\mu} (\bs\;')} \exp\left(\beta J\bs_1\cdot\bs\;'\right)\int \mycolor{\dd \tilde{\mu} (\bs\;'')} \exp\left(\beta J\bs_2\cdot\bs\;''\right)} = \frac{B\exp\left[\beta J\left(\bs_1\cdot\vec{t}_1+\bs_2\cdot \vec{t}_2\right) \right]}{(4 \pi)^2 I_{0,3}^2(\beta J)},
\end{equation}
\mycolor{where $I_{0,3}(\beta J)$ are the generalized modified Bessel functions introduced at the beginning of section 6 of \cite{coolen2005finitely}, restricted to eigenfunctions in the form of equation (72) of \cite{coolen2005finitely}}, 
\begin{equation}\label{eq:kernel}
    \int \mycolor{\dd \tilde{\mu} (\vec{t}_1)} \mycolor{\dd \tilde{\mu} (\vec{t}_2)} \; \psi(\vec{t}_1, \vec{t}_2)=0, 
\end{equation}
is equal to unity. \mycolor{Notice that, differently from (62), there is no average over the disorder (the matrix $\mathbf{U}$ of (62)), because one can integrate out the random sign of the couplings once these kind of eigenfuncitons are taken into account.}
Here $B$ is the branching ratio, $J$ is the absolute value of couplings and \mycolor{$\title{\mu}$ is the uniform measure over the unit sphere (and not over the sphere of radius $\sqrt{3}$, as $\mu$) in dimension \mycolor{$m=3$}}. In our case, $B=Z-1$ since we consider a random regular graph and $J=1/\sqrt{Z-1}$; moreover, since our spins have norm $\sqrt{3}$, we shall rescale $\beta\rightarrow 3\beta$ in eq. \eqref{eq:kernel_coolen}.

In the following, we show how to derive \eqref{eq:equation_for_Tg}.
The function 
\begin{equation}
    \psi^*(\vec{t}_1, \vec{t}_2)\,=\,\frac{1}{2}\vec{t}_1\cdot\vec{t}_2
\end{equation}
is an eigenfunction of \eqref{eq:kernel_coolen}. In fact, one has that ($\beta'\,=\,3\beta/\sqrt{Z-1}$)
\bea
    & \int\mycolor{\dd \tilde{\mu} (\vec{t}_1)} \mycolor{\dd \tilde{\mu} (\vec{t}_2)} \; \mathcal{K}(\bs_1, \bs_2, \vec{t}_1, \vec{t}_2)\frac{1}{2}\vec{t}_1\cdot\vec{t}_2\,=\,\frac{Z-1}{2}
    \frac{\partial \log \int\mycolor{\dd \tilde{\mu} (\vec{t}_1)} e^{\beta' \bs_1\cdot\vec{t}_1}}{\partial (\beta' \bs_1)}\cdot\frac{\partial \log \int \mycolor{\dd \tilde{\mu} (\vec{t}_1)} e^{\beta' \bs_2\cdot\vec{t}_2}}{\partial (\beta' \bs_2)} \\
    & \nonumber \\
    & \,=\,(Z-1)\frac{f_3(\beta')^2}{2}\bs_1\cdot\bs_2 \nonumber
\eea
where $f_3(\beta')$ is the function defined in \eqref{eq:langevin_function}. Imposing that the eigenvalue defined in this last equation is equal to one, we find the critical condition \eqref{eq:equation_for_Tg}. We verified that $\lambda_*\,=\,(Z-1)f_3(\beta')^2$ is the largest odd eigenvalue numerically, by approximating the kernel \eqref{eq:kernel_coolen} with a $500\times 500$ matrix and diagonalizing it.

\mycolor{For completeness, we show that~\eqref{eq:kernel_coolen} follow straightforwardly from~\eqref{eq:finPert} by considering zero external field and the the paramagnetic solution uniform on the sphere, $\hat{\nu}_{i \to j} = \frac{1}{4 \pi}$. Indeed, in this case, ~\eqref{eq:finPert} becomes
\begin{equation}
    \delta \hat{\nu}_{j \to i}(\vec{s}_i) = \frac{1}{\hat{\mathcal{Z}}_{j \to i}}\Big [\int \text{d} \tilde{\mu}(\vec{s}_j) \; e^{\beta J_{ij}\vec{s}_i\cdot \vec{s}_j} \sum_{k \in  \partial j \backslash i} \delta \hat{\nu}_{k \to j}(\vec{s}_j) \left ( \frac{1}{4 \pi}\right)^{B-1} - \frac{\delta \hat{\mathcal{Z}}_{j \to i}}{4 \pi} \Big ]
\end{equation}
where
\begin{equation}
    \hat{\mathcal{Z}}_{j \to i} = \left ( \frac{1}{4 \pi}\right)^{B} \int \dd \tilde{\mu} (\vec{t}_1) \dd \tilde{\mu} (\vec{t}_2) \;  e^{\beta J \vec{t}_1 \cdot \vec{t}_2} = \left ( \frac{1}{4 \pi}\right)^{B-2} I_{0,3}(\beta J) = \hat{\mathcal{Z}}
\end{equation}
where the sign of $J_{ij}$ can be integrated out, thus leaving an edge-independent normalization factor $\hat{\mathcal{Z}}$.
Now we consider $\< \delta \hat{\nu}_{j \to i}(\vec{s}_1) \delta \hat{\nu}_{j \to i}(\vec{s}_2) \>$, where the average is performed over all the pair of $i$ and $j$ corresponding to links in the system. Using the fact that the average perturbation is zero, and requiring that $\< \delta \hat{\nu}_{k \to j}(\vec{s}_1) \delta \hat{\nu}_{\ell \to j}(\vec{s}_2) \> = 0$ when the average is performed on all $j$ and $k \neq \ell$, one finds (for clarity, we drop the subscripts)
\begin{equation}
 \< \delta \hat{\nu}(\vec{s}_1) \delta \hat{\nu}(\vec{s}_2) \> =  \frac{1}{\hat{\mathcal{Z}}^2}\Big [B \int \text{d} \tilde{\mu}(\vec{t}_1)\int \text{d} \tilde{\mu}(\vec{t}_2) \; e^{\beta J_{ij} (\vec{s}_1\cdot \vec{t}_1+\vec{s}_2\cdot \vec{t}_2)} \< \delta \hat{\nu}(\vec{t}_1) \delta \hat{\nu}(\vec{t}_2) \> \left ( \frac{1}{4 \pi}\right)^{2(B-1)} + \< \left ( \frac{\delta \hat{\mathcal{Z}}_{j \to i}}{4 \pi} \right )^2 \> \Big ].
\end{equation}
The second term at the second member of the previous equation is a constant that is equal to zero under the additional requirement that
\begin{equation}
    \int \dd \tilde{\mu} (\vec{t}_1) \dd \tilde{\mu} (\vec{t}_2) \; \< \delta \hat{\nu}(\vec{t}_1) \delta \hat{\nu}(\vec{t}_2) \>=0,
\end{equation}
which corresponds to \eqref{eq:kernel}. Then one finds that 
\begin{equation}
 \< \delta \hat{\nu}(\vec{s}_1) \delta \hat{\nu}(\vec{s}_2) \> =  \frac{B}{ (4 \pi)^2 I_{0,3}^2(\beta J)}\int \text{d} \tilde{\mu}(\vec{t}_1)\int \text{d} \tilde{\mu}(\vec{t}_2) \; e^{\beta J_{ij} (\vec{s}_1\cdot \vec{t}_1+\vec{s}_2\cdot \vec{t}_2)} \< \delta \hat{\nu}(\vec{t}_1) \delta \hat{\nu}(\vec{t}_2) \>,
\end{equation}
which is exactly an eigenvalue problem with kernel \eqref{eq:kernel_coolen}. Therefore  the factor $\lambda$ of the main text coincides with the largest eigenvalue of~\eqref{eq:kernel_coolen}.
}

\end{document}